\documentclass[fleqn,usenatbib]{mnras}

\usepackage{newtxtext,newtxmath}
\usepackage[T1]{fontenc}

\usepackage{graphicx}	
\usepackage{amsmath}	
\usepackage{xcolor}
\usepackage[flushleft]{threeparttable}
\usepackage{verbatim}
\usepackage{array}
\usepackage{tabularx}
\usepackage{colortbl}
\usepackage{hyperref}
\usepackage{float}
\usepackage{subcaption}
\usepackage{booktabs}
\usepackage{adjustbox}
\usepackage{pdflscape}
\usepackage{longtable}

\DeclareRobustCommand{\VAN}[3]{#2}
\let\VANthebibliography\thebibliography
\def\thebibliography{\DeclareRobustCommand{\VAN}[3]{##3}\VANthebibliography}

\newcommand{\msun}{\mbox{$M_{\odot}$}}
\newcommand{\rsun}{\mbox{$R_{\odot}$}}

\newcommand{\teff}{$T_{\rm eff}$}
\newcommand{\feh}{\mbox{$\rm [Fe/H]$}}

\newcommand{\logg}{$\log g$}

\title[The NCORES Program]{The NCORES Program: Precise planetary masses, null results, and insight into the planet mass distribution near the radius gap\thanks{Based on observations collected at the European Southern Observatory under ESO programme 1102.C-0249 (PI: Armstrong). This paper includes data gathered with the 6.5 meter Magellan Telescopes located at Las Campanas Observatory, Chile.}}

\author[D. J. Armstrong et al.]{\parbox{\textwidth}{\Large
David~J.~Armstrong$^{1,2}$, 
Ares~Osborn$^{3}$,
Remo Burn$^{4}$,
Julia Venturini$^{5}$,
Vardan Adibekyan$^{6}$,
Andrea Bonfanti$^{7}$,
Jennifer A. Burt$^{8}$,
Karen A. Collins$^{9}$,  
Elisa Delgado Mena$^{10,6}$,
Andreas Hadjigeorghiou$^{1,2}$,
Steve Howell$^{11}$,
Sam Quinn$^{9}$,  
Sergio G. Sousa$^{6}$,
Marcelo Aron F. Keniger$^{1,2}$,
David Barrado$^{10}$,
S.~C.~C.~Barros$^{6}$,
Daniel Bayliss$^{1,2}$,
Fran\c{c}ois Bouchy$^{5}$, 
Amadeo Castro-González$^{10}$,
Kevin I. Collins$^{12}$, 
Denis M. Conti$^{13}$, 
Ian M. Crossfield$^{14}$, 
Rodrigo Diaz$^{15}$, 
Xavier Dumusque$^{5}$, 
Fabo Feng$^{16,17}$, 
Kathryn V. Lester$^{11}$, 
Jorge Lillo-Box$^{10}$,
Rachel A. Matson$^{18}$, 
Elisabeth C. Matthews$^{19}$,
Christoph Mordasini$^{20}$,
Felipe Murgas$^{21,22}$, 
Hugh P. Osborn$^{20}$, 
Enric Palle$^{21,22}$,
Nuno Santos$^{6,23}$, 
Richard P. Schwarz$^{9}$,  
Tom\'{a}s Azevedo Silva$^{6}$, 
Keivan Stassun$^{24}$, 
Paul Str\o m$^{1,2}$,
Thiam-Guan Tan$^{25}$, 
Johanna Teske$^{26}$,
Gavin Wang$^{27}$,
Peter~J.~Wheatley$^{1,2}$
\\
\footnotesize{Author affiliations can be found at the end of the text.}
}}

\date{\vspace{-0.5cm}Accepted XXX. Received YYY; in original form ZZZ}

\pubyear{2024}

\begin{document}
\label{firstpage}
\pagerange{\pageref{firstpage}--\pageref{lastpage}}
\maketitle


\begin{abstract}
NCORES was a large observing program on the ESO HARPS spectrograph, dedicated to measuring the masses of Neptune-like and smaller transiting planets discovered by the \textit{TESS} satellite using the radial velocity technique. This paper presents an overview of the programme, its scientific goals and published results, covering 35 planets in 18 planetary systems. We present spectrally derived stellar characterisation and mass constraints for five additional TOIs where radial velocity observations found only marginally significant signals (TOI-510.01, $M_p = 1.08^{+0.58}_{-0.55}M_\oplus$), or found no signal (TOIs 271.01, 641.01, 697.01 and 745.01). A newly detected non-transiting radial velocity candidate is presented orbiting TOI-510 on a 10.0d orbit, with a minimum mass of $4.82^{+1.29}_{-1.26}M_\oplus$, although uncertainties on the system architecture and true orbital period remain. Combining the NCORES sample with archival known planets we investigate the distribution of planet masses and compositions around and below the radius gap, finding that the population of planets below the gap is consistent with a rocky composition and ranges up to a sharp cut-off at $10M_\oplus$. We compare the observed distribution to models of pebble- and planetesimal-driven formation and evolution, finding good broad agreement with both models while highlighting interesting areas of potential discrepancy. Increased numbers of precisely measured planet masses in this parameter space are required to distinguish between pebble and planetesimal accretion.
\end{abstract}




\section{Introduction} \label{sec:intro}
The \textit{TESS} mission \citep{Ricker:2014fy} was launched in 2018 with the aim of searching for transiting exoplanets in a new, all-sky survey. The increased sky coverage of the mission over past efforts such as \textit{Kepler} \citep{Borucki:2010dn} led to a substantial increase in the number of planets detected, particularly at Neptune and larger sizes \citep{Guerrero2021}. The brighter candidate host stars observed by an all-sky survey mean that radial velocity observations to confirm planetary mass and density are possible on a much wider scale than previously.

Many programs exist to follow-up \textit{TESS} transiting planet candidates, and that follow-up can take several forms. Photometric follow-up is used to detect the transit with alternative instruments \citep{Collins:2019a}, confirming the nature of the signal, improving ephemeris determination and reducing the possibility that the transit occurs on a background star. High spatial resolution imaging is used to further identify unresolved stars blended with the target in the \textit{TESS} pixels \citep{LBox:2024a,Michel:2024a}. Spectrographs are used to identify spectroscopic binaries and brown dwarfs, characterise the host stars, and, with enough precision, determine planetary masses \citep[e.g.][]{Turtelboom:2022a,Vines:2023a}. Obtaining the planetary mass, along with radius from the transit observation, reveals the planetary density and allows us to build an understanding of possible internal structure, composition, and enable atmospheric measurements, all valuable information in determining the past formation and evolution of planetary systems.

The various forms of follow-up for the \textit{TESS} mission are often coordinated by the \textit{TESS} Follow-up Observing Program \citep[TFOP, ][]{Collins:2018a}, within which the NCORES program described in Section \ref{sect:ncores} operates. Several TFOP-generated observations are published in this paper and the referenced publications. The coordination aims at reducing duplicated observations and enabling broader coverage of planetary candidates, and has helped generate a level and uniformity of follow-up coverage for TESS candidates rarely seen in previous transit surveys.

One particular characteristic of the small planet distribution is the radius gap \citep{Fulton:2017bp, VanEylen:2018ab, Fulton:2018ab}, an observed bimodal distribution in planet radius separating rocky worlds and `sub-Neptunes'. The gap is a predicted outcome of several models, arising either from thermal mass loss after disc dispersal \citep{Lopez:2013ab,Owen:2017kf,Ginzburg:2018ab, Gupta:2019ab,Gupta:2022ab}, differences in the formation process \citep{Lopez:2018ab, Lee21}, or a combination of the two \citep{Venturini:2020rvalley, Venturini:2024, Burn:2024}. These models are dependent on initial assumptions about the planetary properties, including planet core mass, envelope composition and mixing, and migration timescales \citep[e.g. ][]{Rogers:2021ab}. Conversely, the location, width, and presence or absence of planets in the gap can lead to constraints on those properties for the wider planet sample \citep{Owen:2017kf,Jin:2018ef,Mordasini:2020ab}. Investigating the mass of planets in and near the gap potentially allows us to constrain the processes that sculpt the radius valley, with \citet{Ho:2024ab} and \citep{Bonfanti24} recently finding evidence for a mixture of scenarios.

\section{The NCORES program}
\label{sect:ncores}
The NCORES program was a dedicated large program on the ESO \textit{HARPS} spectrograph aimed at studying planetary cores in detail, targeting their composition, formation, and mass distribution. The program targeted transiting planet candidates from \textit{K2} and \textit{TESS}, with radii approximately between 1.8 and 4$R_\oplus$, and orbital periods less than 10 days. The radius limits were chosen to take advantage of the precision of \textit{HARPS} while allowing a substantial sample of targets to be observed. Candidates outside this range were occasionally included due to being in multiple systems with an in-range planet, or through particular scientific interest. The goal was to find planets exposed to intense irradiation where the outer atmosphere may have been evaporated, reducing the degeneracy in identifying the composition and internal structure of the remaining material. Additional goals included characterising planets on either side of the radius gap, noting that the radius gap rises to larger radii for the most irradiated planets \citep{Fulton:2017bp}, to investigate theories of how the gap relates to the core and composition of planets, and building a sample of small planets with precisely characterised masses from the \textit{TESS} mission.

In total, 72 nights on the \textit{HARPS} spectrograph were awarded over two years, in periods P102--P105. Due to operational closures associated with the Covid-19 pandemic, additional time was granted and the program continued through P107, overall covering October 2018 to September 2021. In this time period, over 1300 radial velocity measurements were taken. With this paper, results on 41 candidates in 23 systems have been published, including 32 precise mass measurements and 35 published planets across 18 planetary systems, several jointly with other programmes, in particular the KESPRINT programme. The complete published output of the programme is summarised in Table \ref{tab:wholeprogram}. As this programme began at the start of the \textit{TESS} mission, a target-of-opportunity strategy was used to select bright, plausible candidates for follow-up. Targets were selected on a rolling basis as TOIs were released, based on passing the vetting checks performed by the \textit{TESS} team and falling within our targeted parameter space. Targets were prioritised based on observability, brightness, and scientific interest, for example multi-planet systems. In later efforts, a more quantifiable merit function will be utilised to improve statistical options with future results (see Section \ref{sect:biases}.)


\begin{table*}
    \caption{Details of targets published with contributions from the NCORES program. Many targets are in collaboration with other programs, particularly the KESPRINT program.}
    \centering
    \begin{tabular}{llllllr}
        \hline
        \hline
        Star & Planet & Orbital Period (d) & Planet Mass $(M_\oplus)$ & Planet Radius $(R_\oplus)$ & V mag$^{a}$ & Publication \\
        \hline
        Published\\
        \hline
        TOI-118 / HD219666 & b & 6.04 & $16.6 \pm 1.3$ & $4.71 \pm 0.17$ & 9.9 & \citet{Esposito:2019} \\
    TOI-125 & b & 4.65 & $9.5 \pm 0.88$ & $2.726 \pm 0.075$ & 11 & \citet{Nielsen:jq} \\
    TOI-125 & c & 9.15 & $6.63 \pm 0.99$ & $2.76 \pm 0.10$ & 11 & \citet{Nielsen:jq} \\
    TOI-125 & d & 19.98 & $13.6 \pm 1.2$ & $2.93 \pm 0.17$ & 11 & \citet{Nielsen:jq} \\
    TOI-141  / HD213885 & b  & 1.008 & $8.83 \pm 0.66$ & $1.745 \pm 0.052$ & 7.9 & \citet{Espinoza:dr}\\
    TOI-141 / HD213885 & c  & 4.79 & $19.95 \pm 1.38$ & -- & 7.9 & \citet{Espinoza:dr}\\
    TOI-220 & b & 10.70 & $13.8 \pm 1.0$ & $3.03 \pm 0.15$ & 10.5 & \citet{Hoyer:fb}\\
    TOI-238   & b & 1.27 & $3.40^{+0.46}_{-0.45}$ & $1.402^{+0.084}_{-0.086}$ & 10.8 & \citet{2024arXiv240204113S}\\ 
    TOI-238   & c & 8.47 & $6.7 \pm 1.1$ & $2.18 \pm 0.18$ & 10.8 & \citet{2024arXiv240204113S}\\ 
    TOI-396 & b & 3.59 & $3.55^{+0.94}_{-0.96}$ & $2.004^{+0.045}_{-0.047}$ & 6.4 & \citet{Bonfanti:2025} \\
    TOI-396 & c & 5.97 & $<3.8$ & $1.979^{+0.054}_{-0.051}$ & 6.4 & \citet{Bonfanti:2025} \\
    TOI-396 & d & 11.23 & $7.1 \pm 1.6$ & $2.001^{+0.063}_{-0.064}$ & 6.4 & \citet{Bonfanti:2025} \\
    TOI-431 & b & 0.49 & $3.07 \pm 0.35$ & $1.28 \pm 0.04$ &  9.1 & \citet{Osborn:fu} \\
    TOI-431 & c & 4.85 & $2.83^{+0.41}_{-0.34}$ & -- & 9.1    & \citet{Osborn:fu} \\
    TOI-431 & d & 12.46 & $9.9 \pm 1.5$ & $3.29 \pm 0.09$ &  9.1   & \citet{Osborn:fu} \\
    TOI-560 / HD73583 & b & 6.40 & $10.2^{+3.4}_{-3.1}$ & $2.79 \pm 0.10$ & 9.7 & \citet{Barragan:ix} \\
    TOI-560 / HD73583 & c & 18.88 & $9.7^{+1.8}_{-1.7}$ & $2.39^{+0.1}_{-0.09}$ & 9.7 & \citet{Barragan:ix} \\
    TOI-755 / HD110113 & b & 2.54 & $4.55 \pm 0.62$ & $2.05 \pm 0.12$ & 10.1 & \citet{Osborn:kw}  \\
    TOI-755 / HD110113 & c & 6.74 & $10.5 \pm 1.2$ & -- & 10.1 & \citet{Osborn:kw}  \\
    TOI-836 & b & 3.82 & $4.53^{+0.92}_{-0.86}$ & $1.70 \pm 0.07$ & 9.9 & \citet{Hawthorn:202349H}\\
    TOI-836 & c & 8.60 & $9.6^{+2.7}_{-2.5}$& $2.59 \pm 0.09$ & 9.9 & \citet{Hawthorn:202349H}\\
    TOI-849 & b & 0.77 & $39.1^{+2.7}_{-2.6}$ & $3.44^{+0.16}_{-0.12}$  & 12.0 & \citet{Armstrong:2020dm}\\
    TOI-969 & b & 1.82 & $9.1 \pm 1.0$ & $2.77^{+0.09}_{-0.10}$ & 11.6 & \citet{LilloBox:202309L} \\
    TOI-969 & c & 1700 & $3590^{+250}_{-290}$ & -- & 11.6 & \citet{LilloBox:202309L} \\
    TOI-1052 & b & 9.14 & $16.9\pm1.7$ & $2.87^{+0.29}_{-0.24}$ & 9.5 & \citet{Armstrong:202304A}\\
    TOI-1052 & c & 35.8 & $34.3^{+4.1}_{-3.7}$ & -- & 9.5 & \citet{Armstrong:202304A}\\
    TOI-1062 & b & 4.11 &  $10.2 \pm 0.8$ & $2.27^{+0.10}_{-0.09}$ & 10.2 & \citet{Otegi:fm} \\
    TOI-1062 & c & 7.97 & $9.78^{+1.26}_{-1.18}$ & -- & 10.2 & \citet{Otegi:fm}\\
    TOI-1097 / HD109833 & b & 9.19 & -- & $2.89^{+0.15}_{-0.13}$ & 9.3 & \citet{Wood:20235W}\\
    TOI-1097 / HD109833 & c & 13.9 & -- & $2.59^{+0.20}_{-0.18}$ & 9.3 & \citet{Wood:20235W}\\
    TOI-1099 / HD207496 & b & 6.44 & $6.1\pm1.6$ & $2.25^{+0.12}_{-0.10}$ & 8.2 & \citet{Barros:20234B}\\
    TOI-2000 & b & 3.10 & $11.0\pm2.4$ & $2.70\pm0.15$ & 11.0 & \citet{Sha:20233S}\\
    TOI-2000 & c & 9.13 & $81.7^{+4.7}_{-4.6}$ & $8.14^{+0.31}_{-0.30}$ & 11.0 & \citet{Sha:20233S}\\
    K2-364 / HD137496 & b & 1.62 & $4.04 \pm 0.55$ & $1.31^{+0.06}_{-0.05}$ & 9.9 & \citet{Silva:2022cs}\\
    K2-364 / HD137496 & c & 479.9 & $2435 \pm 35$ & -- & 9.9 & \citet{Silva:2022cs}\\
 \hline 
    \end{tabular}
    
    $^{a}$\,From the \textit{TESS} Input Catalogue v8 \citep{Stassun2019TIC}. Where no radius is given, masses are minimum masses.
    \label{tab:wholeprogram}
\end{table*}

\section{Observations} \label{sec:obs}
Planets with confirmed mass measurements have been published in various papers listed in Table \ref{tab:wholeprogram}. Readers should see those papers for details of data used and modelling for each case. Five additional targets where the planetary candidate could not be convincingly detected are published here, to ensure the programme's null results are in the public domain, to allow other teams to use the mass constraints to inform their targeting, and to enable these constraints to be incorporated into mass-radius relations to avoid biases towards high mass planets at small radii. These null result targets, namely TOI-271, TOI-510, TOI-641, TOI-697 and TOI-745, along with their identifiers are listed in Table \ref{tab:starcoords}. Additional observations undertaken by TFOP on the five new targets are detailed in this section. TOI-880 (Nielsen et al in prep) and TOI-1839 (Castro-González et al in prep) were also observed, but continued to receive multiple seasons of data after the NCORES programme finished, and will be published separately.

\begin{table*}
    \centering
    \caption{Stellar identifiers for new targets in this paper.}

    \begin{tabular}{llllllr}
        \hline
        \hline
        Star & Alt. ID & TIC ID$^{a}$ & Gaia ID$^{b}$ & R.A. [h:m:s] & Dec. [d:m:s]  & TESS mag$^{a}$  \\
        \hline
       TOI-271 & HIP 21952 & 259511357 & 4784439709132505344 & 04:43:06.12 & $-50$:13:10.12 &  $8.433 \pm 0.007$  \\
       TOI-510 & HIP 33681 & 238086647 & 5508330367833229824 & 06:59:50.16 & $-49$:30:26.93 &  $8.463 \pm 0.006$  \\
       TOI-641& -- & 49079670 & 2925175888849429888 & 06:44:17.21 & $-23$:49:09.07 &  $9.904 \pm 0.006$ \\
       TOI-697 & -- & 77253676 & 4866555051425383424 & 04:38:49.01 & $-36$:40:52.77&  $9.355 \pm 0.006$   \\
       TOI-745 & -- & 444842193 & 5307756013606927616 & 09:51:01.78 & $-55$:19:05.88   & $10.549 \pm 0.006$ \\
        \hline
    \end{tabular}
    
     $^{a}$\,From the \textit{TESS} Input Catalogue v8 \citep{Stassun2019TIC}.$^{b}$\,From Gaia DR3 \citep{Gaia2023}.
    \label{tab:starcoords}
\end{table*}

\subsection{\textit{TESS} photometry}
Each additional candidate planet in this paper was first identified using \textit{TESS} photometry. \textit{TESS} observes in $\sim$27-day long sectors, and each target has at the time of writing been observed in multiple, different sectors, listed in Table \ref{tab:models} along with cadence and radial velocity data obtained. All targets were observed at a 2-minute cadence, except TOI-641 which has a mixture of 30-minute and 2-minute sectors.

\begin{table*}
    \caption{Model setup for new targets in this paper. TOI-510 includes an additional non-transiting candidate.}

    \centering
    
    \begin{tabular}{llllllllr}
        \hline
        \hline
        Star & Period (d) & \textit{TESS} sectors & Cadence (min) & N. \textit{HARPS} & N. other RV & GP (Photometry)  & RV Linear drift & Fixed $e$,$\omega$\\
        \hline
       TOI-271 & 2.48 & 3,4,5,30,31,32 & 2 & 15 & 94 (PFS) & No  & No & Yes\\
       TOI-510 & 1.35, 10.0 & 6,7,8,33,34,35,61,62 & 2 & 36 & 0 & Yes & Yes & Yes \\
       TOI-641 & 1.89 & 6,7,33 & 30 (S6,S7), 2 (S33) & 9 & 6 (PFS) & No  & No & Yes \\
       TOI-697 & 8.61 & 4,5,31,32 & 2 & 6 & 0 & Yes  & No & Yes \\
       TOI-745 & 1.08 & 9,10,36,37,63 & 2 & 26 & 0 & Yes  & No & Yes \\
        \hline
    \end{tabular}
    \label{tab:models}
\end{table*}

TOIs 271.01, 510.01, 641.01 and 697.01 were identified by the \textit{TESS} Science Processing Operations Center pipeline \citep[SPOC pipeline;][]{Jenkins-16,Jenkins:2010b,Jenkins:2002}. TOI-745.01 was identified by the MIT Quick Look Pipeline \citep[QLP, ][]{Huang:2020qlp2}. Although the TOI detection was performed on various fractions of the currently available dataset, our joint model considers the full data available at the time of writing, listed in Table \ref{tab:models}.

For the purpose of modelling the data to provide mass constraints, we utilised the \textit{TESS} data provided by the SPOC pipeline for all targets, which performs light curve detrending via Presearch Data Conditioning Simple Aperture Photometry (PDCSAP) \citep{Stumpe:2014ep,Stumpe:2012bj,Smith:2012ji}. Variability is seen in the lightcurves of TOI-510, TOI-697 and TOI-745. Notably TOI-510 shows quasiperiodic variability at twice the candidate period. For these targets the variability is later accounted for using Gaussian processes during the joint fit. Data from separate sectors were stitched together by dividing the sector by its median flux value to normalise each sector independently. 

\begin{figure}
    \centering
    \includegraphics[width=\columnwidth]{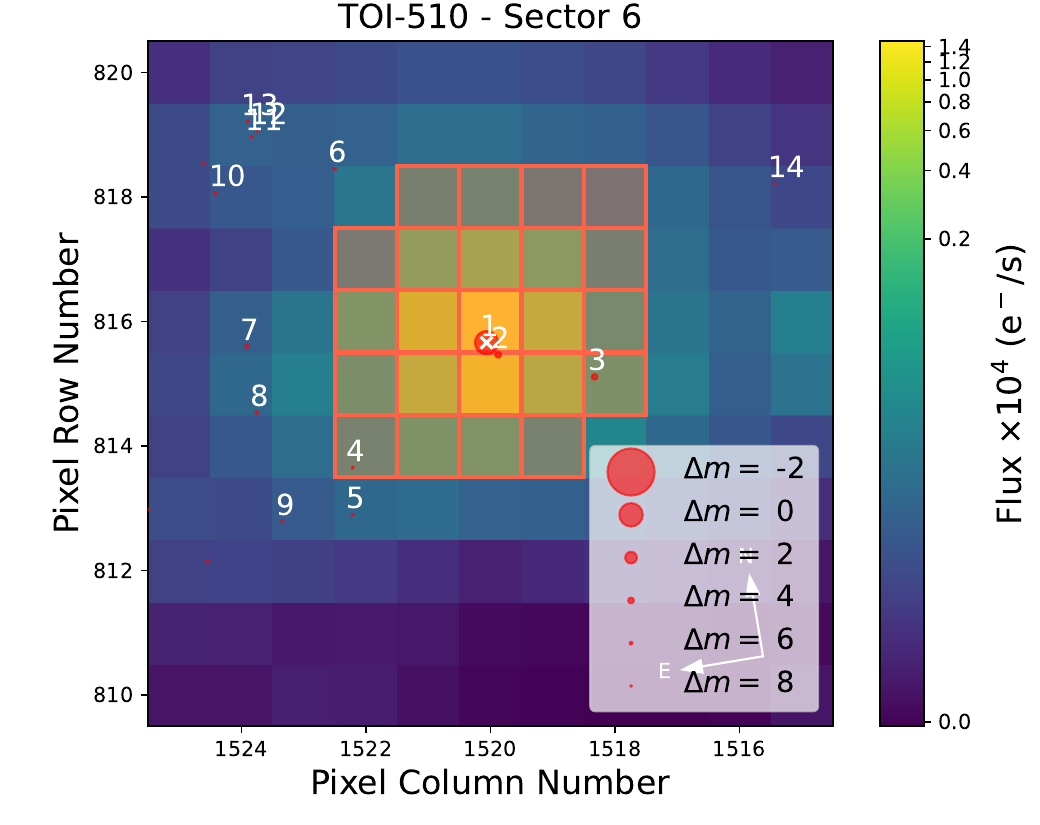}
    \caption{\textit{TESS} field of view around TOI-510 in Sector 6, generated with \texttt{tpfplotter}. Known stars from Gaia are shown as red circles. TOI-510 is marked by a white cross. The \textit{TESS} photometric aperture is highlighted in red.}
    \label{fig:tpf510}
\end{figure}

\subsection{Light Curve Follow-up}

The \textit{TESS} field of view near TOI-510 can be seen in Figure \ref{fig:tpf510}, plotted using the \texttt{tpfplotter} package \citep{Allertpfplotter}\footnote{www.github.com/jlillo/tpfplotter)}. The field of view of TOIs-271, 641, 697 and 745 are shown in the Appendix. In each case some nearby stars can be seen, motivating further follow-up. We acquired ground-based time-series follow-up photometry of the stars in the fields around all targets in Table \ref{tab:starcoords} as part of the \textit{TESS} Follow-up Observing Program \cite{Collins:2019a}\footnote{https://tess.mit.edu/followup} to attempt to rule out or identify nearby eclipsing binaries (NEBs) as potential sources of \textit{TESS} detections at the times of transit predicted by the public SPOC and QLP ephemerides. We used the {\tt TESS Transit Finder}, which is a customized version of the {\tt Tapir} software package \citep{Jensen:2013}, to schedule our transit observations. The observations, facilities, and results are summarized in Table \ref{table:transitnebfollowup}. The Las Cumbres Observatory Global Telescope \citep[LCOGT;][]{Brown:2013} images were calibrated by the standard LCOGT {\tt BANZAI} pipeline \citep{McCully:2018}. The LCOGT and SLR2 photometric data were extracted using {\tt AstroImageJ} \citep{Collins:2017}. The Perth Exoplanet Survey Telescope used a custom pipeline based on {\tt C-Munipack}\footnote{http://c-munipack.sourceforge.net} to calibrate the images and extract the differential photometry. The MuSCAT2 \citep{Narita:2019} and MEarth-South \citep{Irwin:2007} results were extracted using the custom pipelines described in \citet{Parviainen:2020} and \cite{Irwin:2007}, respectively.

Since the TOI depths are generally shallower than can typically be reliably detected with ground-based follow-up observations, we initially checked for possible NEBs that could be contaminating the SPOC and QLP photometric apertures, which generally extend $\sim1\arcmin$ from the target star. To account for possible contamination from the wings of neighboring star PSFs, we searched for NEBs in all known Gaia EDR3 and TIC version 8 nearby stars out to $2\farcm5$ from the target stars that are bright enough in \textit{TESS} band (assuming a 100\% eclipse) to produce the \textit{TESS} detection. To attempt to account for possible delta-magnitude differences between \textit{TESS}-band and the follow-up filter band, we included stars that are up to 0.5 magnitudes fainter. We consider a star cleared of an NEB within the above mentioned sensitivity limits if the RMS of its 10-minute binned light curve is more than a factor of 3 smaller than the expected NEB depth in the star. We then visually inspect each neighboring star's light curve to ensure no obvious eclipse-like signal. All of our follow-up data are available on the {\tt EXOFOP-TESS} website\footnote{https://exofop.ipac.caltech.edu/tess}.



\begin{table*}
\begin{threeparttable}
 \centering
 \caption{Ground-based light curve observations.
 }
 \label{table:transitnebfollowup}
    \begin{tabularx}{\linewidth}{@{\extracolsep{\fill}}lclccl} 
    \toprule
    \toprule
Observatory  & Aper (m) & Location                 & UTC Date      & Filter              &    Result\\
    \midrule
TOI-271.01 & & & & &\\
    \cline{1-1}
PEST\tnote{1}          & 0.3  & Perth, Australia          & 2018-12-23   &  $\rm R_c$           &  cleared all possible NEBs\tnote{2} \\
LCOGT\tnote{3}-CTIO    & 1.0  & Cerro Tololo, Chile       & 2018-12-26   &  Sloan $r'$          &  cleared all possible NEBs \\
[1.5mm]TOI-510.01 & & & & &\\
\cline{1-1}
PEST                   & 0.3  & Perth, Australia          & 2019-03-28   &  $\rm R_c$         &  no obvious NEBs \\
LCOGT-SAAO             & 1.0  & Cape Town, South Africa   & 2019-03-29   &  Sloan $i'$   &  cleared NEBs, except $6\arcsec$ neighbor\\
SLR2-SAAO\tnote{4}     & 0.5  & Cape Town, South Africa   & 2019-03-29   &  $\rm I_c$    &  cleared NEBs, except $6\arcsec$ neighbor\\
[1.5mm]TOI-641.01 & & & & &\\
\cline{1-1}
MEarth-South           & $0.4\times7$ & Cerro Tololo, Chile & 2019--11-06 & RG715        &  no obvious NEBs\\
[1.5mm]TOI-697.01 & & & & &\\
\cline{1-1}
PEST                   & 0.3  & Perth, Australia          & 2019-10-06   &  $\rm R_c$           &  cleared all possible NEBs \\
[1.5mm]TOI-745.01 & & & & &\\
\cline{1-1}
LCOGT-CTIO             & 1.0  & Cerro Tololo, Chile       & 2021-01-16   &  $z$-short\tnote{5}  &  cleared NEBs in 60\% of 852 neighbors\\
\bottomrule
\bottomrule
\end{tabularx}

\begin{tablenotes}
       \item [1] Perth Exoplanet Survey Telescope
       \item [2] See text for a definition of stars cleared of NEBs 
       \item [3] Las Cumbres Observatory Global Telescope \citep{Brown:2013} 
       \item [4] {\it Solaris} network of telescopes of the Nicolaus Copernicus Astronomical Center of the Polish Academy of Sciences.
       \item [5] Pan-STARRS $z$-short 
     \end{tablenotes}
  \end{threeparttable}
\end{table*}


\subsection{High resolution imaging}
\label{sec:hr}






If an exoplanet host star has a close companion, that companion (bound or chance-aligned) can create a false-positive transit signal if it is, for example, an eclipsing binary (EB), or lead to underestimates of the exoplanet radius and incorrect stellar and planet properties \citep{Lillo-Box2014aa, Ciardi2015, Furlan2017, Furlan2020}. Searching for nearby stellar companions is an especially important check to make for main sequence stars with large planets, due to the possible confusion between a large planet and a stellar eclipse \citep{Lillo-Box2021aa}. The discovery of close, bound companion stars, which exist in nearly one-half of FGK type stars \citep{Matson2018} and just under one-quarter of M class stars \citep{Susemiehl2022}, provides crucial information toward our understanding of exoplanet formation, dynamics and evolution \citep{Howell2021AJ}. Close companions can even cause non-detections of small planets residing within the same exoplanetary system \citep{Lester2021}. Thus, to search for close-in bound companions unresolved in \textit{TESS} or other typical ground-based follow-up observations, we obtained high-resolution imaging speckle observations of the new targets.

\begin{figure}
    \centering
    \includegraphics[width=\columnwidth]{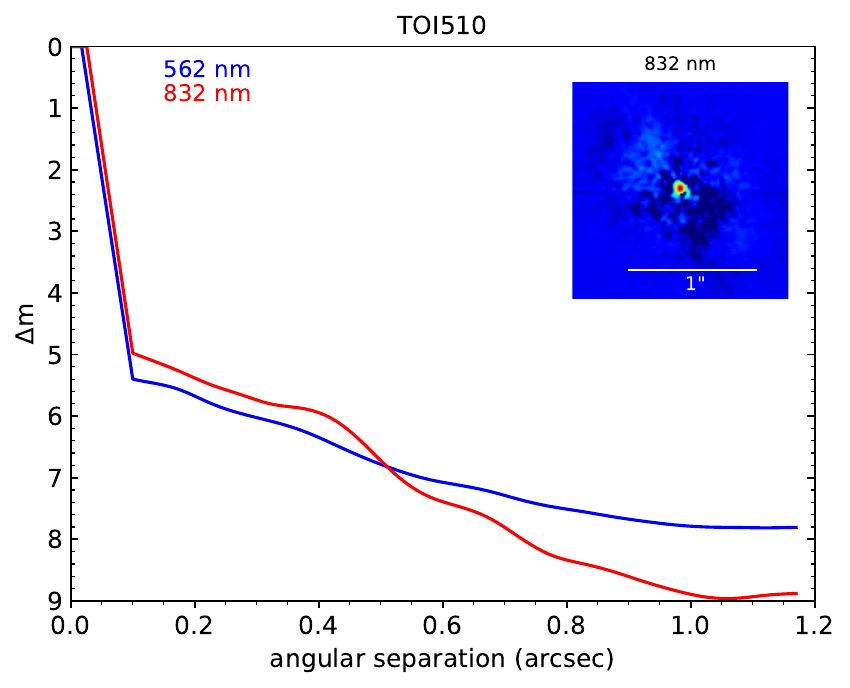}
    \caption{Contrast curves for TOI-510 obtained by Gemini/Zorro at 562 and 832nm. No companion star is detected within the contrast limits.}
    \label{fig:gemini}
\end{figure}

These TOIs were observed with the Gemini North or Gemini South telescope between September 2019 and September 2021\footnote{The reduced data and precise dates of observation for each star are publicly available at the NASA Exoplanet Archive: https://exoplanetarchive.ipac.caltech.edu/}  using the speckle interferometric instruments ‘Alopeke and Zorro \citep{Scott2021}. The two instruments are duplicates of each other and both provide simultaneous speckle imaging in two bands (562nm and 832 nm) with output data products including a reconstructed image with robust contrast limits on companion detections \cite[e.g. ][]{Howell2016}. Three to five sets of $1000\times0.06$ sec exposures were collected for these stars as described in \citet{Howell2022} and subjected to Fourier analysis in the standard reduction pipeline \citep[see ][]{Howell2011}. Figure \ref{fig:gemini} shows an example of final contrast curves and an 832 nm reconstructed speckle image, typical of what we obtained for each TOI. We find that TOI-271, TOI-641, and TOI-697 reveal one or more close companion stars and they have been taken to be binary or triple stars \citep[see][]{Lester2021}. Individual cases are discussed in Section \ref{sec:results}. TOI-510 and TOI-745 are single stars to within the angular (0.02 to 1.2 arcsec) and magnitude contrast limits (5-7 magnitudes) achieved. Given that all of these TOIs are about 100 pc away, the angular limits above correspond to spatial limits of roughly 2 AU to 120 AU at each star.

We also observed TOI-641 with NIRI at Gemini South on UT 2019-11-03 using the observatory's adaptive optics system.  We obtained 9 dithered images in the Br-$\gamma$ filter with exposure times of 3.8~s to avoid saturating the primary star. The raw data products are available at the Gemini Data Archive; the processed products are at the ExoFOP website. The stacked observations reach a contrast level of $\sim$7.5~mag beyond $\sim$0.4'' and reveal two companions, with separation, position angle, and delta-magnitudes of 3.537'', 86.2$^\circ$, and 4.16$\pm$0.03~mag and 3.627'', 80.7$^\circ$, and $3.90\pm0.02$~mag, respectively.


\subsection{HARPS}

\textit{HARPS} \citep[High Accuracy Radial velocity Planet Searcher;][]{Mayor-03} is an Echelle spectrograph mounted on the ESO 3.6\,m telescope located at La Silla Observatory, Chile. In total, around 1300 spectra were taken as part of the NCORES program (PI Armstrong, ID 1102.C-0249) throughout periods P102--P107, 1st October 2018 to 31st September 2021. The majority of these spectra are published in other works listed in Table \ref{tab:wholeprogram}. Newly published analyses in this work are for TOI-271 (15 spectra, 900-1200\,s exposure time), TOI-510 (36 spectra, 900\,s), TOI-641 (9 spectra, 1800\,s), TOI-697 (6 spectra, 1800\,s), and TOI-745 (26 spectra, 1800\,s). Data were obtained in \textit{HARPS} High-Accuracy Mode with a 1\arcsec\ diameter fibre and standard resolution of R$\sim$115,000. Raw data were reduced according to the standard \textit{HARPS} data reduction software, with radial velocities extracted using a cross-correlation function with a G2 spectral template \citep[][]{Pepe-02,Baranne-96}. The data table for these observations can be found in Table~\ref{tab:spec1}, which we use in our joint modelling (Section~\ref{sect:jointmod}).

\subsection{PFS}

TOI-271 and TOI-641 were additionally observed by the \textit{PFS} spectrograph. \textit{PFS} \citep[the Carnegie Planet Finder Spectrograph;][]{Crane2006,Crane2008,Crane2010} is a custom designed precision RV echelle spectrometer covering 3900 to 6700 \AA\ on the 6.5m Magellan Clay telescope at Las Campanas Observatory, Chile. With the exception of the focus, PFS has no moving parts and is embedded in an insulated box where the temperature is maintained at 27$^{\circ}\pm$ 0.01$^{\circ}$C. As of January 2018, PFS uses a 10K $\times$ 10K detector with 9$\mu$m pixels which produces a peak resolution of $R \simeq$ 130,000 when using the 0.3" slit and default 1x2 binning, as was done for all observations of TOI-271 and TOI-641. PFS RVs are measured in the 5000 - 6200 \AA\ region of the spectrum where an Iodine cell placed in the converging beam of the telescope provides a dense forest of sharp absorption lines that enable precise wavelength calibration \citep{MarcyButler1992}. The PFS Iodine cell was scanned with the NIST Fourier Transform Spectrometer \citep{Nave2017} at a resolution of 1,000,000. Raw PFS data is reduced to 1D, wavelength calibrated, spectra with a custom built raw reduction package and velocities are generated from an updated version of the Iodine-based, forward modelling approach outlined in \citet{Butler1996}. TOI-271 was observed 94 times by PFS, with a consistent exposure time of 600 seconds. TOI-641 was observed 6 times. The data table for these observations can be found in Table~\ref{tab:specpfs}, which we use in our joint modelling.

\section{Stellar parameters and chemical abundances}            \label{sec:parameters}

\begin{table*}
\begin{threeparttable}
    \caption{Stellar parameters for new targets in this paper.}
    \label{tab:starpars}
    \centering
    \begin{tabular}{llllllllr} 
        \hline
        \hline
        Star & $R_*$ [\rsun{}]& $M_s$ [\msun{}] &  \teff{} [K] & \logg{}  & \feh{}& $v_\mathrm{tur}$  & Age $[{\rm Gyr}]$ & Age  $[{\rm Gyr}]$ \\
        & &  &   &   & &  & Isochrones & Chem. clocks$^\star$ \\
        \hline
       TOI-271 & $1.274 \pm 0.038$ & $1.147 \pm 0.027$ & $6139 \pm 65$ & $4.17 \pm 0.11$ & $0.01 \pm 0.04$  & $1.313\pm0.024$ & $3.7 \pm 1.0$ & $4.41\pm0.14$\\
       TOI-510 & $1.259 \pm 0.040$ & $0.878 \pm 0.010$ & $5800 \pm 63$ & $4.27 \pm 0.11$ & $-0.48 \pm 0.04$  & $1.171\pm0.028$ & $11.3 \pm 0.3\dagger$ & $>12$\\
        & &  &  & &   &  & $12.6^{+0.8}_{-2.5}\dagger$ & \\
       TOI-641& $1.002 \pm 0.029$ & $0.982 \pm 0.034$ & $5607 \pm 62$ & $4.38 \pm 0.12$ & $0.12 \pm 0.04$  & $0.917\pm0.02$ & $5.6 \pm 3.1$ & $6.04\pm0.63$\\
       TOI-697 &  $0.926 \pm 0.021$ & $0.967 \pm 0.033$ & $5561 \pm 64$ & $4.47 \pm 0.11$ & $0.10 \pm 0.04$  & $0.874\pm0.026$ & $3.5 \pm 2.9$ & $3.84\pm0.38$\\
       TOI-745 & $0.996 \pm 0.022$ & $0.912 \pm 0.028$ & $5737 \pm 61$ & $4.38 \pm 0.11$ & $-0.17 \pm 0.04$  & $0.97\pm0.016$ & $8.4 \pm 2.3$ & $10.19\pm0.56$\\
        \hline
\end{tabular}
\begin{tablenotes}
       \item $\star$ Errors are the weighted mean errors of different chemical clocks. A value of 1.5Gyr is a more realistic age error for the method (see text).
       \item $\dagger$ See text for description of different age estimation techniques
     \end{tablenotes}
  \end{threeparttable}

\end{table*}

\begin{table*}
    \caption{Chemical abundances for new targets in this paper.}

    \centering
    \begin{tabular}{lllllllr}
        \hline
        \hline
          Star	&	[Mg/H]&		[Al/H]&		[Si/H]	&	[Ti/H]&		[Ni/H]   &  [C/H] & [Cu/H]   \\
        \hline
    TOI-271	 &		$-0.01\pm0.05$ &		$-0.06\pm0.07$ &		$0.01\pm0.01$ &		$-0.02\pm0.03$ &		$-0.02\pm0.02$ &  $0.044\pm0.027$ & $-0.048\pm0.041$ \\
    TOI-510	&		$-0.20\pm0.05$ &	$-0.26\pm0.03$ &		$-0.29\pm0.02$ &		$-0.21\pm0.03$ & 		$-0.45\pm0.02$ &   $-0.257\pm0.039$ & $-0.497\pm0.056$  \\
    TOI-641	&		$0.13\pm0.03$ &	$0.17\pm0.04$ &		$0.12\pm0.02$ &		$0.15\pm0.02$ &		$0.13\pm0.02$  &     $0.091\pm0.033$ & $0.183\pm0.053$ \\ 
    TOI-697	&	$0.10\pm0.04$ &	$0.11\pm0.04$ &		$0.08\pm0.02$ &		$0.14\pm0.03$ &		$0.09\pm0.02$   & $-0.010\pm0.014$  & $0.080\pm0.027$ \\
    TOI-745	&		$-0.12\pm0.04$ &	$	-0.13\pm0.05$ &		$-0.15\pm0.02$ &		$-0.13\pm0.02$ &	$-0.19\pm0.01$ &  $-0.190\pm0.020$ & $-0.173\pm0.029$  \\
        \hline
    Star & [Zn/H]  & [Sr/H]  &  [Y/H]  & [Zr/H]  & [Ba/H] & [Ce/H] &[Nd/H]   \\
    \hline 
    TOI-271  &   $0.015\pm0.024$  &  $-0.021\pm0.077$  &  $-0.004\pm0.039$  &  $-0.100\pm0.105$  &   $0.005\pm0.050$  &  $-0.084\pm0.072$  &  $-0.040\pm0.033$  \\
     TOI-510 &  $-0.314\pm0.037$  &  $-0.533\pm0.077$  &  $-0.547\pm0.058$  &  $-0.444\pm0.026$  &  $-0.611\pm0.042$  &  $-0.575\pm0.073$  &  $-0.463\pm0.032$ \\
     TOI-641   &   $0.123\pm0.016$  &   $0.053\pm0.077$  &   $0.046\pm0.154$  &   $0.070\pm0.031$  &  $0.031\pm0.026$  &   $0.072\pm0.024$  &   $0.131\pm0.021$ \\
     TOI-697   &   $0.041\pm0.010$  &   $0.180\pm0.077$ &   $0.198\pm0.132$ &   $0.143\pm0.058$ &   $0.107\pm0.031$ &   $0.183\pm0.053$  &   $0.173\pm0.020$ \\
     TOI-745  &  $-0.120\pm0.022$  &  $-0.229\pm0.077$ &  $-0.190\pm0.149$ &  $-0.204\pm0.043$ &  $-0.201\pm0.025$ &  $-0.180\pm0.062$  &  $-0.165\pm0.041$ \\
     \hline
    \end{tabular}
    \label{tab:starchems}
\end{table*}

We co-added the individual exposures for each target in order to create a combined HARPS master spectrum with a higher signal-to-noise ratio. We used each master spectrum to derive the stellar spectroscopic parameters ($T_{\mathrm{eff}}$, $\log g$, microturbulence, [Fe/H]) and the respective uncertainties following the ARES+MOOG methodology as described in \citet[][]{Sousa-21, Sousa-14, Santos-13}. The ARES code \footnote{The last version of ARES code (ARES v2) can be downloaded at https://github.com/sousasag/ARES} \citep{Sousa-07, Sousa-15} was used to measure in a consistent way the equivalent widths (EW) of iron lines included in the line list presented in \citet[][]{Sousa-08}. Briefly ARES+MOOG performs a minimization process looking for the ionization and excitation equilibrium to find convergence on the best set of spectroscopic parameters. For the computation of the iron abundances we make use of a grid of Kurucz model atmospheres \citep{Kurucz-93} and the radiative transfer code MOOG \citep[v2019;][]{Sneden-73}. Following the same methodology as described in \citet[][]{Sousa-21}, we used the distance derived from the GAIA eDR3 parallaxes and estimated the trigonometric surface gravity for each target. We then estimated the stellar age, mass and radius using the derived spectroscopic parameters with the \texttt{PARAMv1.3} tool \citep{2006A&A...458..609D, 2014MNRAS.445.2758R, 2017MNRAS.467.1433R} using the PARSEC isochrones \citep{2012MNRAS.427..127B}.

Under the assumption of local thermodynamic equilibrium, the classical curve-of-growth analysis technique was used. The stellar abundances of the elements were also determined using the same methods and models used to determine the stellar parameters. Although the EWs of the spectral lines were measured automatically with ARES, for some elements (Al, Mg, C, O) with only two-three lines available, we performed careful visual inspection of the EWs measurements. For the derivation of chemical abundances of refractory elements we closely followed the methods described in \citet[e.g.][]{Adibekyan-12, Adibekyan-15, DelgadoMena-17} whereas for the derivation of C and O abundances we followed the methods described in \citet{DelgadoMena-21, Bertrandelis-15}. The results of our analysis are shown in Tables \ref{tab:starpars} and \ref{tab:starchems}. In a subsequent step, we used the derived abundances to obtain estimations of the stellar age by applying calibrations based on chemical clocks, i.e. certain abundance ratios with a strong correlation with age, such as [Y/Mg], [Y/Zn] or [Sr/Si] \citep[see Table 10 in ][with the 3D formulas employed here]{DelgadoMena2019}. The derived ages (weighted mean of different clocks) are in agreement with the values from \texttt{PARAMv1.3}. We note that the errors reported in the table are the weighted mean errors of the different clocks and their low values show the agreement between the different chemical clocks. A value of 1.5 Gyr should be considered as a more realistic age error for this method.

The chemical pattern of the observed stars suggest that, excepting TOI-510, all the stars belong to the Galactic thin disk. TOI-510 shows enhancement in $\alpha$ elements (Mg, Si, Ti), which is typical for the chemically defined thick disk population \citep[e.g.][]{Adibekyan-11, Adibekyan-13, Lagarde-21}. TOI-510 also appears to be a somewhat older star than the others in the sample, consistent with thick disk membership.

To confirm TOI-510's old evolutionary stage, we also followed another statistical treatment, by applying the isochrone placement routine \citep{bonfanti2015,bonfanti2016} that interpolates the input parameters along with their uncertainties (i.e. $T_{\mathrm{eff}}$, [Fe/H], and $\log{g}$) within pre-computed grids of PARSEC\footnote{\url{http://stev.oapd.inaf.it/cgi-bin/cmd}} v1.2S \citep{marigo2017} isochrones and tracks. We further derived $\log{R'_{HK}}=-4.95\pm0.02$ from the HARPS spectra to improve the interpolation scheme convergence as detailed in \citet{bonfanti2016,bonfanti2020} and obtained an age of $12.5_{-2.5}^{+1.3}$ Gyr.
Aware that the spectroscopic $\log{g}$ we adopted as input in the previous derivation process might be poorly constrained, we performed other isochrone placement runs by replacing the input $\log{g}$ with the Gaia parallax (offset-corrected following \citet{lindegren2021}) and different magnitude bands. In other words, the isochrone interpolation was done in the absolute magnitude-$\log{T_{\mathrm{eff}}}$ diagram, rather than in the $\log{g}$-$\log{T_{\mathrm{eff}}}$ (also known as the Kiel diagram). In detail, we performed one isochrone placement run per magnitude (namely the Gaia $G$, $G_{\mathrm{BP}}$, and $G_{\mathrm{RP}}$, the Johnson $B$ and $V$, the Tycho $B_T$ and $V_T$, and the 2MASS $J$, $H$, and $K_s$). From each run we derived a probability density function (pdf) for the age and finally we merged (i.e. we summed) the pdfs after checking their mutual consistency according to the $\chi^2$-based criterion outlined in \citet{bonfanti2021}. We obtained an age of $12.6_{-2.5}^{+0.8}$ Gyr, which is consistent with the outcome from our first approach, the chemical clocks, and the initial PARAM isochrone age, confirming the old evolutionary stage of TOI-510.

\section{Joint Models}
\label{sect:jointmod}
The modelling applied to published cases is dependent on the particular system and described in the relevant papers listed in Table \ref{tab:wholeprogram}. Here we describe the joint model applied in this work to new systems.

We model the data for each target using the \texttt{exoplanet} \citep{exoplanet:joss} code framework. This package also makes use of \texttt{starry} \citep{starryLuger2019} and \texttt{PYMC} version 5 \citep{SalvatierPymc3}.

We restrict ourselves to using \textit{TESS} data in the joint model; although some other lightcurves are available from ground-based instruments, these are not of good enough precision to contribute to the transit fit. The whole \textit{TESS} lightcurve is used.

The flux is normalised to zero for all of the photometry by dividing the individual light curves by the median of their out-of-transit points and subtracting one. To model the planetary transits, we used a limb-darkened transit model following the \citet{kipping2013} quadratic limb-darkening parameterisation, and Keplerian orbit models. Relevant parameters are the stellar radius $R_*$ in solar radii, the stellar mass $M_*$ in solar masses, the orbital period $P$ in days, the planetary radii $R_p$, the time of a reference transit $t_0$, the impact parameter $b$, the eccentricity $e$, and the argument of periastron $\omega$. One Keplerian orbit is initialised for each planet in a system. 

In all cases $e$ and $\omega$ were consistent with zero for a planet in initial model runs, and we fix $e$ and $\omega$ to 0 in the final joint fit model. We use values from the \textit{TESS} pipelines to initialise our priors on the epochs, periods, transit depths and radii of the transiting planets, and the results of our stellar characterisation to set priors on the stellar mass and radius. Prior distributions are given in Table \ref{tab:priors}. For the non-transiting TOI-510 c a wider Normal distribution prior on period and epoch was used, with mean $\mu$ of 10.5d and 1167.7 (BJD-2458000) and standard deviation $\sigma$ of 4d and 10d. The prior for $P_c$ was bounded between 5 and 25 days to ensure stability in the fitting chains.

Typically, a simple jitter term was added in quadrature to the \textit{TESS} data to account for additional noise. For TOI-510, TOI-697 and TOI-745, the \textit{TESS} light curves show additional correlated stellar variability. The correlated variability was modelled with the SHOTerm Gaussian Process (GP) given in \textsc{exoplanet} \footnote{\url{https://celerite2.readthedocs.io/en/latest/api/python/\#celerite2.terms.SHOTerm}}, representing a stochastically-driven, damped harmonic oscillator. We set this up using the hyperparameters $\log{(sigma)}$, $\log{(rho)}$, and $\tau$. The prior on $\log{(sigma)}$ (where $sigma$ represents the standard deviation of the process) was set to a normal distribution with a mean equal to the log of the standard deviation of the \textit{TESS} flux and a standard deviation of 3. The prior on $\log{(rho)}$ (where $rho$ represents $2\pi/\omega_0$, with $\omega_0$ being the frequency of the undamped oscillator) was also set as a normal distribution but with mean 0 and standard deviation 3. $\tau$ represents the damping timescale of the process and was fixed to 2 days. We then take the sum of our model planetary light curves and subtract these from the total \textit{TESS} flux. The GP is trained on the resulting residual light curve simultaneously to remove the stellar variability.

We also fit separate systematic offset and jitter terms for each radial velocity instrument, and, in the case of TOI-510, a linear drift term. The means of the jitter priors are set to the log of the minimum error on the relevant dataset. No GP is used on any of the radial velocity data, as the activity indicators available from HARPS do not show clear signs of periodic variability. Note that given the relatively small number of HARPS points in some cases, particularly TOI-271, TOI-641 and TOI-697, this should not be taken as definitive information on the activity levels of the targets.

To fit the model, we use the \verb|PyMC| sampler, which draws samples from the posterior using a variant of Hamiltonian Monte Carlo, the No-U-Turn Sampler (NUTS). We allow for 1500 burn-in samples which are discarded, followed by 8250 steps for each of 8 chains. 

Table \ref{tab:models} details the specific model setup for each system.

\begin{table}
    \small
    \caption{Prior distributions used for transiting planets in our joint fit model. The priors are created using distributions in \texttt{PyMC} with the relevant inputs to each distribution described in the table footer.}
	\label{tab:priors}
	\begin{threeparttable}
	\begin{tabular}{l p{1.3cm} l}
	\toprule
	\textbf{Parameter} & \textbf{(unit)} & \textbf{Prior Distribution}  \\
	\midrule
	Period $P$                      & (days)               & $\mathcal{N}$(\textit{TESS} period, 0.001)       \\
	Ephemeris $t_{0}$                 & (BJD)        & $\mathcal{N}$(\textit{TESS} epoch, 0.04)   \\
	Radius $\log{(R_p)}$                   & (log \mbox{R$_{\odot}$}) & $\mathcal{N}(*, 1.0)$       \\
        Impact Parameter $b_b$         &                    &       $\mathcal{U}(0, 1+R_p/R_{*})$         \\
         $e_b\sin\omega_b$                &                    & 0 (fixed)                            \\
	$e_b\cos\omega_b$                   &         & 0 (fixed)                            \\
	$K_b$                     & (m\,s$^{-1}$)        & $\mathcal{U}(0.0, 50.0)$              \\
	\midrule
	\multicolumn{3}{l}{\textbf{Star}} \\
	Mass $M_{*}$                    & (\mbox{M$_{\odot}$}) & $\mathcal{N_B}$(Table \ref{tab:starpars} $\mu$ and $\sigma$, 0.5, 2.0)  \\
	Radius $R_{*}$                  & (\mbox{R$_{\odot}$}) & $\mathcal{N_B}$(Table \ref{tab:starpars} $\mu$ and $\sigma$, 0.5, 2.0)   \\
	\midrule	
	\multicolumn{3}{l}{\textbf{Photometry}} \\
	\textit{TESS} mean              &        Rel. Flux            & $\mathcal{N}(0.0, 1.0)$                \\
	$\log{(\rm{Jitter)}}$          &        & $\mathcal{N}({\dagger}, 3.0)$                \\
    $\log{(\rm{sigma)}}$ (GP only)  &   & $\mathcal{N}({\dagger}, 3.0)$ \\
     $\log{(\rm{rho)}}$ (GP only)  &  & $\mathcal{N}(0.0, 3.0)$   \\
        \midrule
	\multicolumn{3}{l}{\textbf{HARPS and PFS RVs}} \\
	Offset                          & (m\,s$^{-1}$)        & $\mathcal{N}(\mu(RV), 20.0)$        \\
	$\log{(\rm{Jitter)}}$           & (log m\,s$^{-1}$)        & $\mathcal{N}({\dagger}, 3.0)$  \\
     Drift (TOI-510 only)               & (m\,s$^{-1}$d$^{-1}$)        & $\mathcal{N}(0.79, 1.0)$        \\
	\bottomrule
	\end{tabular}
	\begin{tablenotes}
	\item \textbf{Distributions:}
	\item $\mathcal{N}(\mu, \sigma)$: a normal distribution with a mean $\mu$ and a standard deviation $\sigma$;
	\item $\mathcal{N_B}(\mu, \sigma, a, b)$: a bounded normal distribution with a mean $\mu$, a standard deviation $\sigma$, a lower bound $a$, and an upper bound $b$ (bounds optional);
	\item $\mathcal{U}(a, b)$: uniform distribution with lower bound $a$, and upper bound $b$.
	\item \textbf{Prior values:}
	\item $*$ $0.5(\log{(D)}) + \log{(R_{*})}$ where $D$ is the transit depth (ppm multiplied by $10^{-6}$) and $R_{*}$ is the mean of the prior on the stellar radius (\mbox{R$_{\odot}$});
	\item ${\dagger}$ the log of the minimum error on the HARPS data (m\,s$^{-1}$), or the standard deviation of the \textit{TESS} flux. We fit a log value to enforce an broad, non-zero prior covering several orders of magnitude.
	\end{tablenotes}
	\end{threeparttable}
\end{table}

\begin{figure}
    \centering
    \includegraphics[width=\columnwidth]{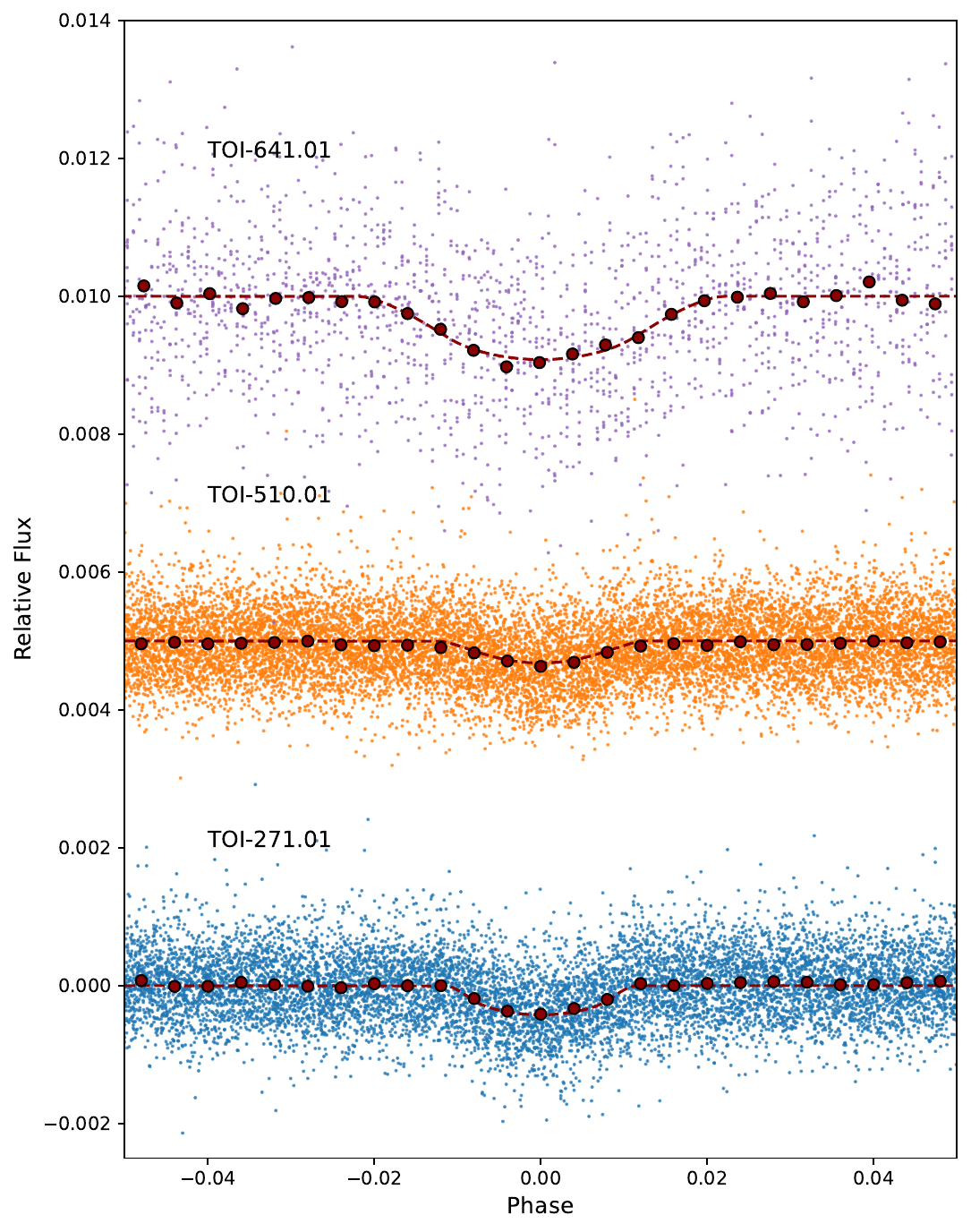}
    \caption{Light curves phase-folded to the best fitting orbital periods and zoomed to the transit region for new candidates considered in this paper. Binned flux is overplotted. Best fitting models are shown as dashed lines. From top to bottom, TOI-641.01, TOI-510.01 and TOI-271.01 are shown.}
    \label{fig:lc271}
\end{figure}

\begin{figure}
    \centering
    \includegraphics[width=\columnwidth]{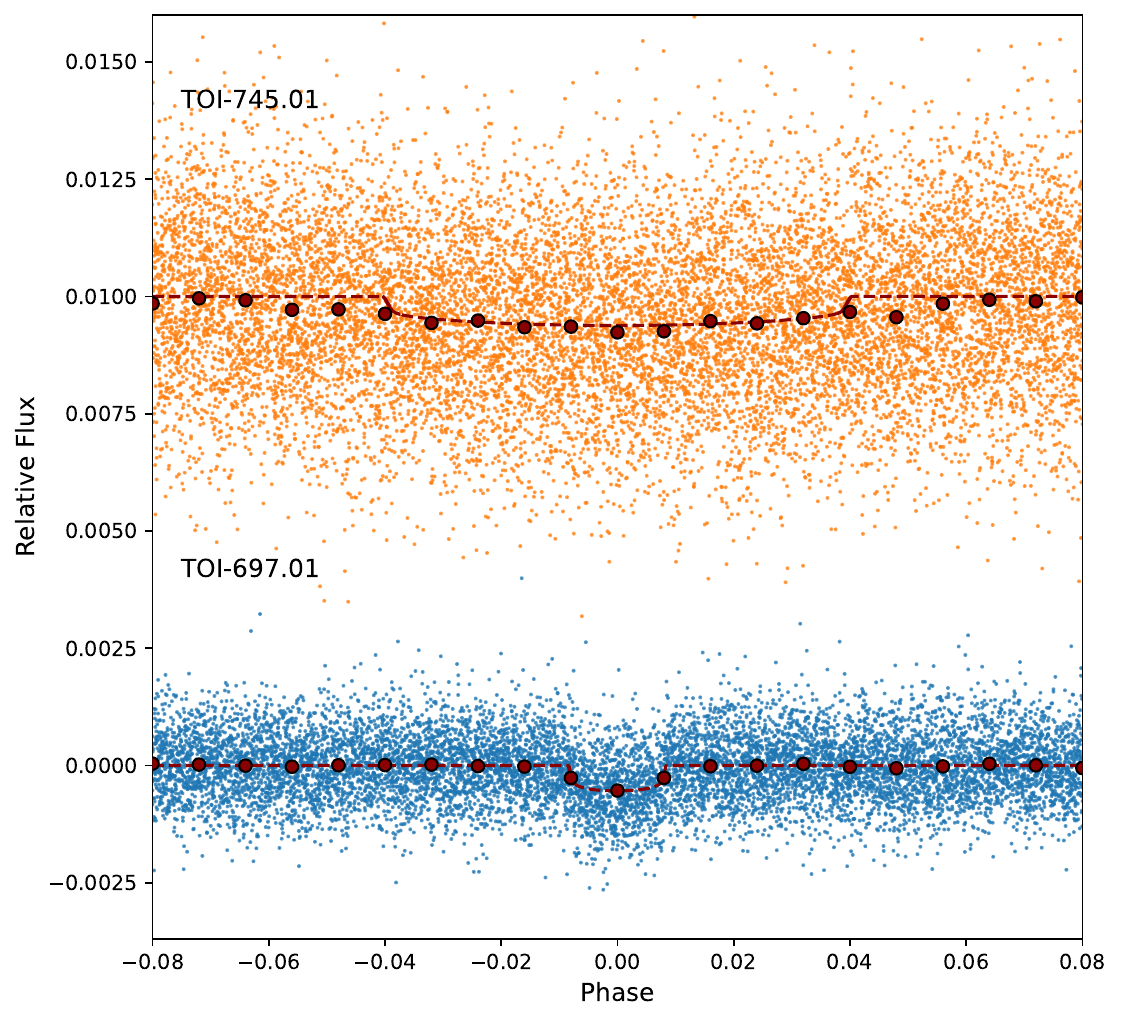}
    \caption{Light curves phase-folded to the best fitting orbital periods and zoomed to the transit region for new candidates considered in this paper. Binned flux is overplotted. Best fitting models are shown as dashed lines. From top to bottom, TOI-745.01 and TOI-697.01 are shown. Some discrepancy in the ingress and egress of TOI-745 can be seen, despite an eccentric model not being preferred.}
    \label{fig:lc697}
\end{figure}

\section{Results} \label{sec:results}

In this section we present the upper limits on the planetary mass from joint models of the new TOIs where radial velocity observations could not detect the candidate planet. The full fit results are given in Tables \ref{tab:instparams}, \ref{tab:fitresults} and \ref{tab:fit_app}. Figures \ref{fig:lc271} and \ref{fig:lc697} show the lightcurves and best fit for each null result.

\subsection{TOI-510/HD52758}
\label{sect:results510}
\begin{figure}
    \centering
    \includegraphics[width=\columnwidth]{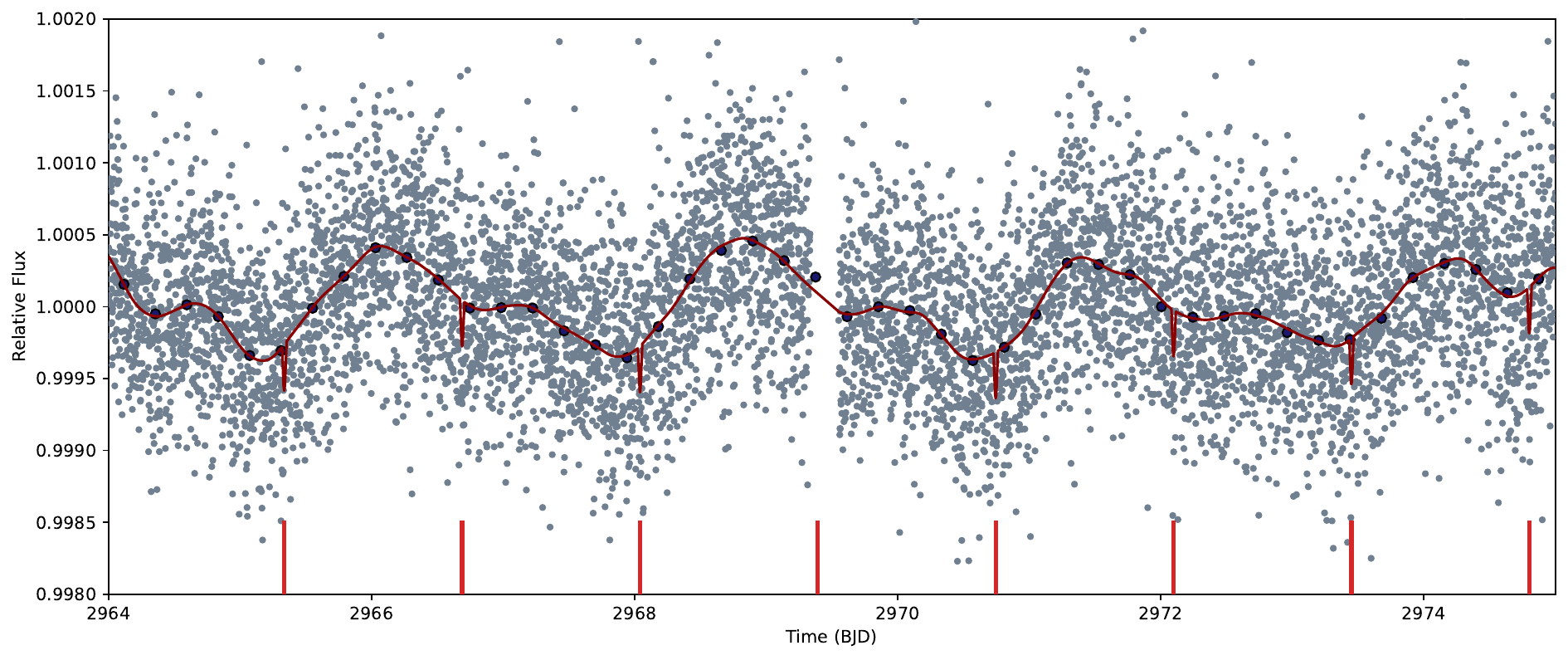}
    \caption{Zoom of the Sector 61 TESS PDCSAP lightcurve for TOI-510, showing correlated variability on twice the transiting candidate period. Transits of TOI-510.01 are marked with vertical lines.}
    \label{fig:toi510_lcvar}
\end{figure}

The TESS lightcurve of TOI-510 shows significant correlated variability at twice the period of the candidate transit (Figure \ref{fig:toi510_lcvar}), which may be connected to stellar rotation, or the orbit of a blended eclipsing binary. Given the old age of the star (Table \ref{tab:starpars}) it is highly unlikely that a 2.7 day periodic signal could be related to the stellar rotation of TOI-510 itself, lending more weight to a blended scenario. Although an apparently significant transit remains in the detrended lightcurve (Figure \ref{fig:lc271}), this transit is V-shaped, implying either a grazing planet, a blended stellar eclipse, or remnant stellar variability. We do not come to a conclusion on the nature of the TOI candidate TOI-510.01 but consider this ambiguity to cast doubt on a planetary origin. The variability has been accounted for using a Gaussian Process in the joint model, which can be seen in Figure \ref{fig:toi510_lcvar}.

\begin{figure}
    \centering
    \includegraphics[width=\columnwidth]{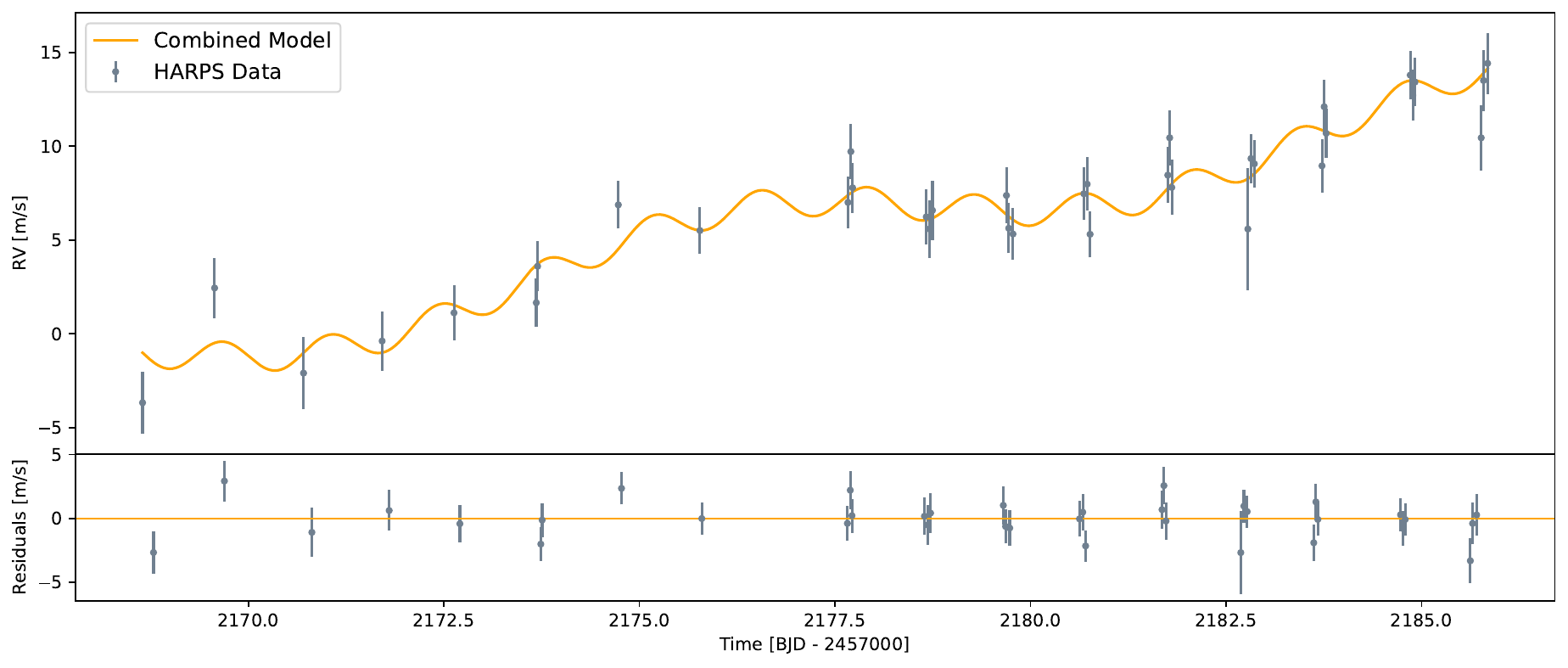}
    \caption{Radial velocity timeseries for TOI-510, showing linear drift. The best fit model is shown in orange.}
    \label{fig:toi510_rvvar}
\end{figure}

\begin{figure}
    \centering
    \includegraphics[width=\columnwidth]{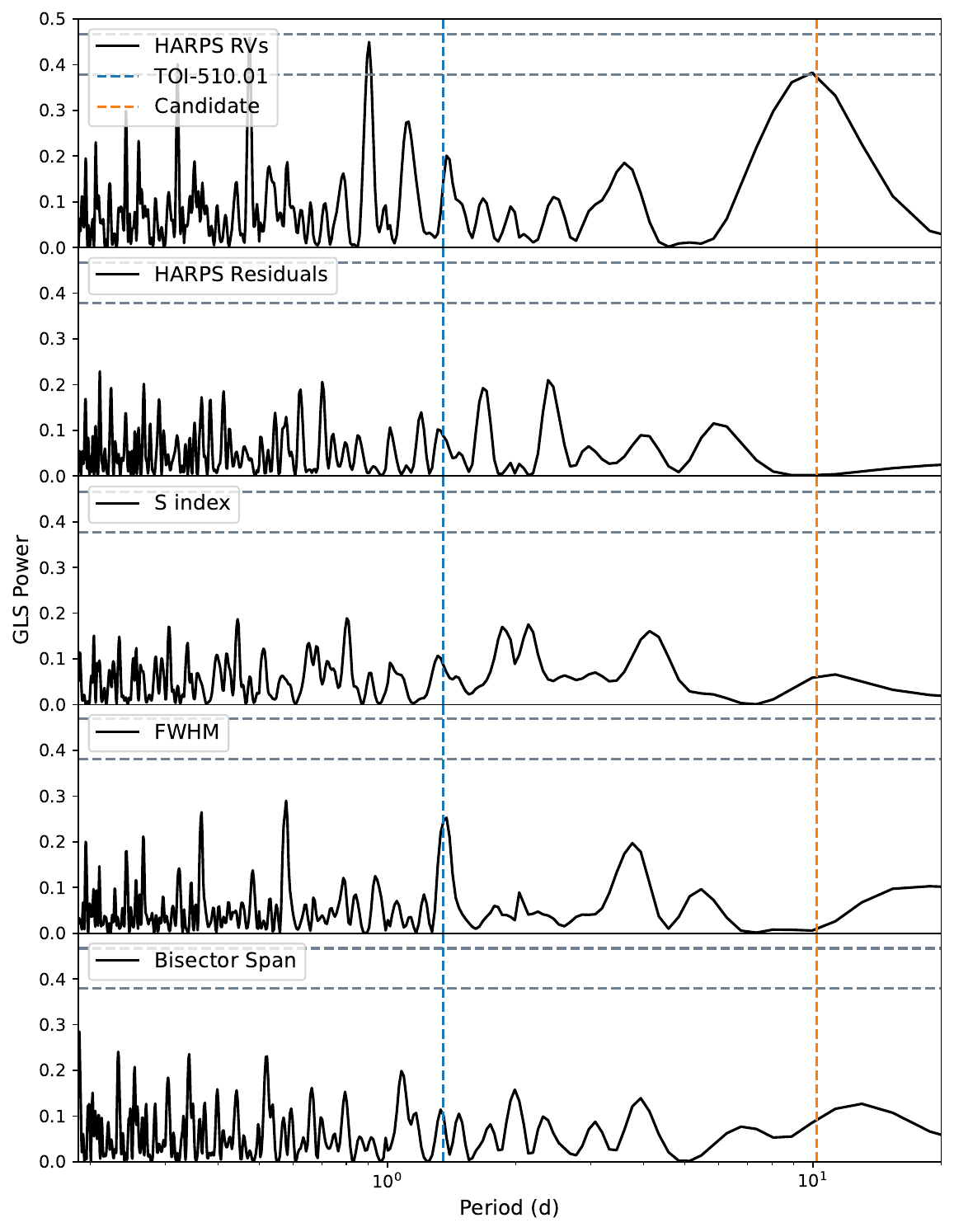}
    \caption{Generalised Lomb-Scargle periodograms for TOI-510 RVs, residuals and activity indicators, after removing the linear trend. False Alarm Probability lines of 0.1 and 0.01 are shown. The transiting candidate period near 1.35d and the non-transiting candidate near 10d are highlighted with vertical dashed lines.}
    \label{fig:toi510_ls}
\end{figure}

In radial velocity, TOI-510 shows a significant linear drift of $0.785^{+0.070}_{-0.065} \mathrm{m\,s}^{-1}\mathrm{d}^{-1}$ (Figure \ref{fig:toi510_rvvar}), a marginal 2$\sigma$ signal $K_b$ at the period of the transiting candidate of $0.66^{+0.35}_{-0.33} \mathrm{m\,s}^{-1}$, and an additional signal on a period of $10.0^{+1.3}_{-1.0}$d with $K=1.56^{+0.40}_{-0.41} \mathrm{m\,s}^{-1}$. The additional 10 day signal is only observed for approximately 2 cycles (Figure \ref{fig:toi510_rvvar}) and only reaches a false alarm probability of 10\%, leading us to consider it as a candidate only. Periodograms of the RV data and activity indicators are shown in Figure \ref{fig:toi510_ls}. Given the short baseline, aliases of the 10d signal are also significant (at approx. 0.9d, 0.47d and 0.32d) and the true period is unclear, requiring more data points to clarify. No significant periodicity is seen in the indicators, although the CCF FWHM does show a spike at the period of the transiting candidate.

\begin{figure}
    \centering
    \includegraphics[width=\columnwidth]{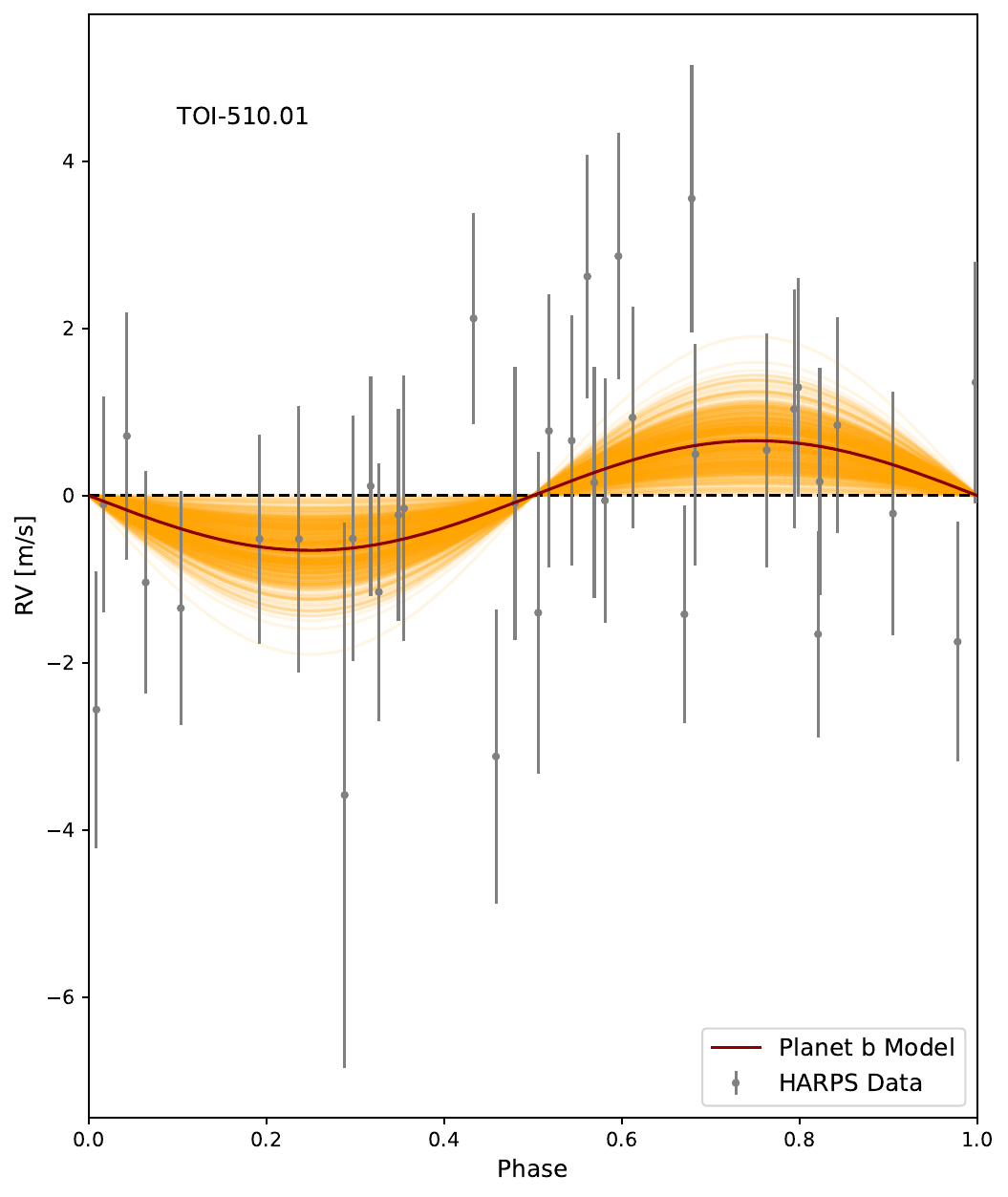}
    \caption{Radial velocity data for TOI-510, phase folded on the period of planet b from the joint model, with the best fitting model of planet c removed. Random models from the MCMC trace are shown in orange. The best fit model is shown in red.}
    \label{fig:toi510b}
\end{figure}

\begin{figure}
    \centering
    \includegraphics[width=\columnwidth]{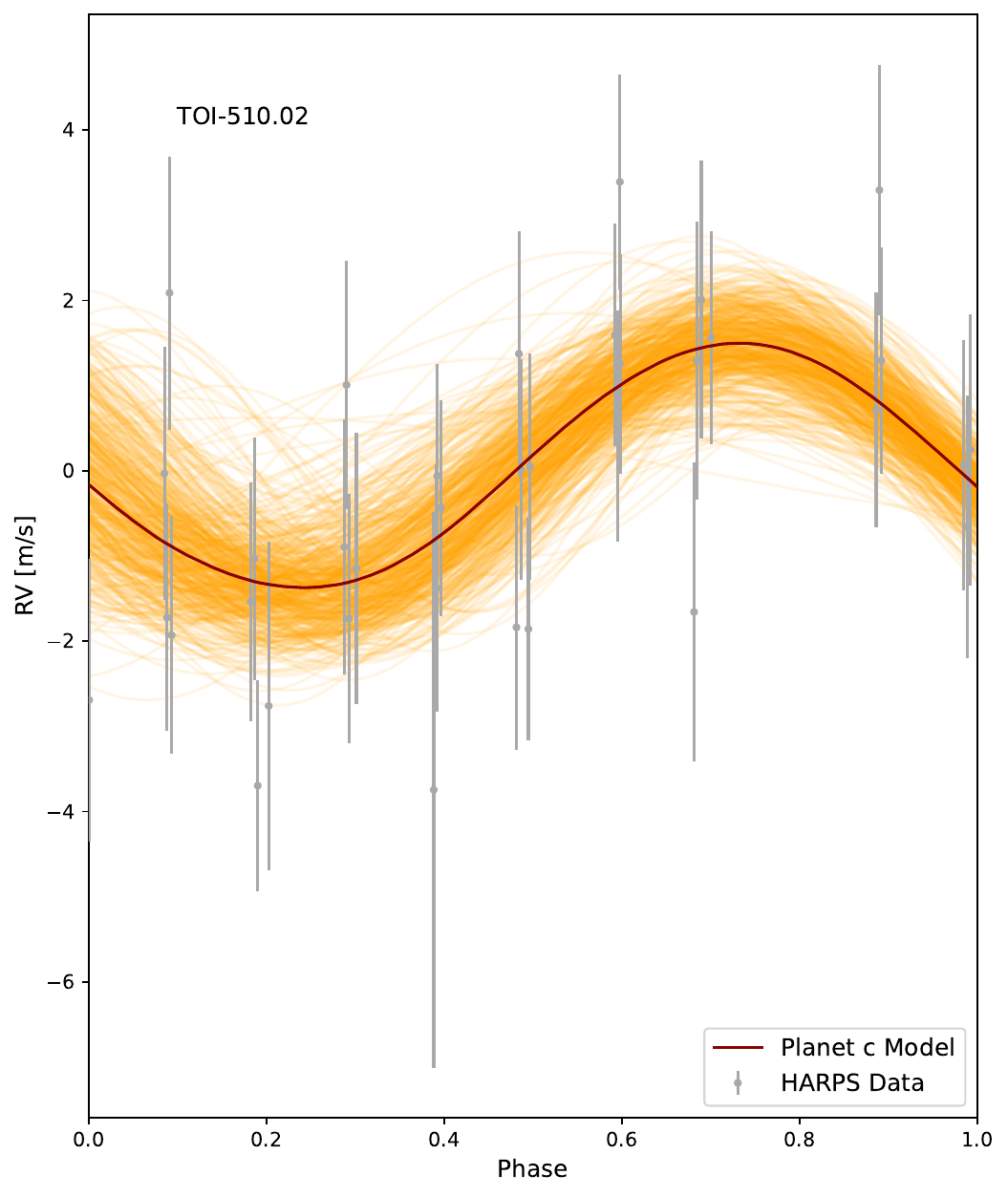}
    \caption{Radial velocity data for TOI-510, phase folded on the period of planet c from the joint model, with the best fitting model of planet b removed. Random models from the MCMC trace are shown in orange. The best fitting model is shown in red.}
    \label{fig:toi510c}
\end{figure}

The corresponding mass of the inner transiting candidate $b$ is $1.08^{+0.58}_{-0.55}M_\oplus$, and $m_c\sin i$ for the non-transiting candidate planet, assuming the 10d peak is correct, is $4.82^{+1.29}_{-1.26}M_\oplus$. The phased RV data and models for each planet are shown in Figures \ref{fig:toi510b} and \ref{fig:toi510c}. The transit fit for planet $b$ is grazing (Figure \ref{fig:lc271}) leading to a poorly constrained radius of approx. $4.18^{+2.03}_{-0.96}R_\oplus$, assuming a planet led to the transiting signal. If real, the planet appears to have a low density compared to other planets of this mass, with $\rho=0.08\mathrm{g\,cm}^{-3}$ using the nominal mass and radius values, but the density is strongly dependent on the poorly constrained radius. No stellar companions are detected by high resolution imaging, although the linear drift implies a larger mass planetary or stellar companion on a wide orbit. Corroborating this, the Gaia RUWE value for TOI-510 is 2.43, implying the star's astrometric motion is inconsistent with a single-star solution. There is also a resolved nearby source in Gaia (TIC238086649) at a separation of 6", visible in Figure \ref{fig:tpf510}. Dilution from this neighbour is removed by the SPOC pipeline, but that removal does not account for variability, and the nearby neighbour could be an eclipsing binary causing the apparent transit, and/or periodic variability. Ground-based transit observations capable of resolving both stars are recommended to clarify the nature of the candidate. We ran the \texttt{TESSPositionalProbability} tool \citep{Hadjigeorghiou2024aa} to determine probabilities for each star to be the true host of the candidate, given the observed centroid. We find a 41\% probability of TOI-510 being the true host, a 46\% probability for TIC238086649, and the remainder on more distant fainter sources. As such the centroid data is inconclusive but does not reject TOI-510 being the true host. Overall this is a complex system which should be treated with caution.


\subsection{Other null results}
\begin{figure*}
    \centering
    \includegraphics[width=\textwidth]{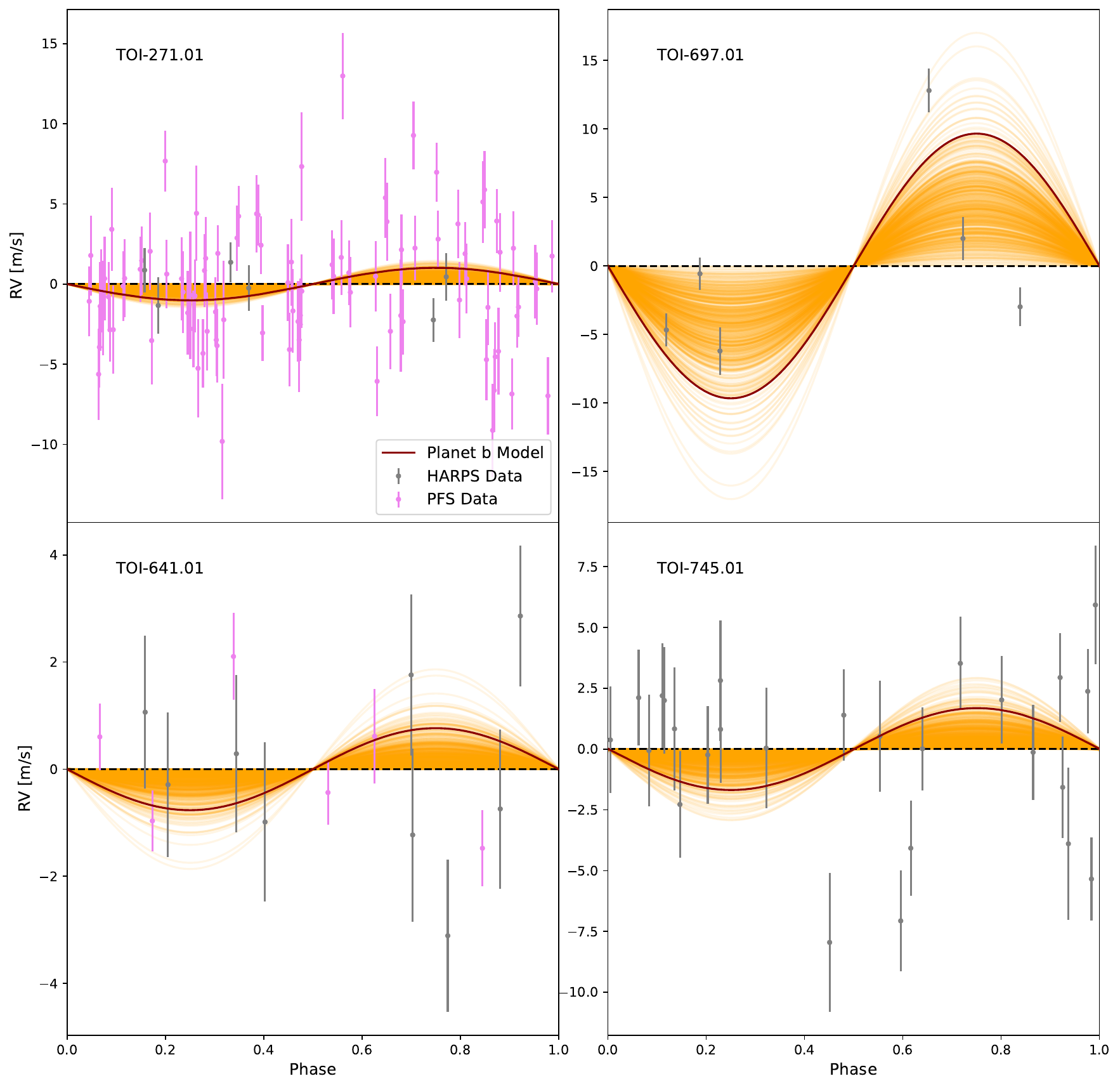}
    \caption{Radial velocity data for TOI-271, TOI-641, TOI-697 and TOI-745, phase folded on the period from the joint model. Random models from the MCMC trace are shown in orange. The model best matching the 95\% confidence limit on semi-amplitude is shown in red to highlight the limit we place on candidate planet mass.}
    \label{fig:rvnulls}
\end{figure*}

\textbf{TOI-271/HIP21952} was observed by both HARPS and PFS. No significant signal is seen in the RVs at the transiting candidate period. Our joint fit finds a 95\% confidence upper limit on $K_b$ of 1.02 $\mathrm{m\,s}^{-1}$, corresponding to an upper limit on planetary mass of $2.39M_\oplus$. The RV data and model are shown in Figure \ref{fig:rvnulls}. The planetary radius is found to be $4.16^{+0.50}_{-0.52}R_\oplus$, but this does not consider dilution from the detected companion star from high spatial resolution imaging, which has a delta magnitude of 5, angular separation 0.15" \citep{Lester2021} and is fully blended with the target star at the resolution of \textit{TESS}, implying 1\% of the flux in the \textit{TESS} aperture comes from the companion star. It is unclear which star the transiting object orbits, and hence what its true radius is accounting for dilution. If the object orbits the primary star, the implied planetary radius is $4.24R_\oplus$, well within the error. If the object orbits the companion star, $R_p/R_*$ becomes 0.301, implying an eclipsing stellar object depending on the unknown parameters of the companion star. Although the TOI disposition remains uncertain, given the stellar companion and null RV detection, this target has a high chance of being a false positive.

\textbf{TOI-641} was observed by both HARPS and PFS. No significant signal is seen in the RVs at the transiting candidate period. Our joint fit finds a 95\% confidence upper limit on $K_b$ of $0.77\mathrm{m\,s}^{-1}$, corresponding to an upper limit on planetary mass of $1.49M_\oplus$. The RV data and model are shown in Figure \ref{fig:rvnulls}. The candidate radius is found to be $3.44^{+0.24}_{-0.22}R_\oplus$ under the planet model, not considering dilution. Two stellar companions are detected by high resolution imaging from Gemini, with delta magnitudes of 3.9 and 4.2 and angular separations of 3.6" and 3.5" respectively. The fraction of flux in the \textit{TESS} aperture from these companions is therefore 4.5\%. Assuming the transiting object transits the primary star, this dilution in the \textit{TESS} lightcurve would imply a planetary radius of $3.60R_\oplus$, large considering the mass upper limit. If the transiting object orbits one of the other bound stars, the radius would be substantially larger and imply a stellar companion. There is additionally a resolved nearby source in Gaia, TIC744547537. We again ran the \texttt{TESSPositionalProbability} tool \citep{Hadjigeorghiou2024aa}, finding a 22\% probability that the star TOI-641 is the true host, and a 67\% probability that TIC744547537 is the source of the observed signal, with the remaining probability on fainter more distant stars visible in Figure \ref{fig:tpf641}. These results combine with the tight upper mass limit to imply that the observed transit likely does not originate from TOI-641.

\textbf{TOI-697} was observed with HARPS, with only 6 data points obtained which show variability out of phase with the transit ephemeris. Performing a simple Keplerian fit gives a 95\% confidence upper limit on $K_b$ of 9.60 $\mathrm{m\,s}^{-1}$, corresponding to an upper limit on planetary mass of $30.1M_\oplus$. The RV data and model are shown in Figure \ref{fig:rvnulls}. The candidate radius is found to be $2.17^{+0.09}_{-0.08} R_\oplus$ under the planet model, not considering dilution. Several stellar companions are detected by high resolution imaging \citep{Lester2021}, implying that TOI-697 is a triple star with two lower mass companion stars ($m_*/m_1=0.41,0.38$). Transit analysis by \citet{Lester-2022} showed that the primary star is the most likely host of the transiting object, with a probability of 87\%, which, if true, means that the impact of dilution on the estimated radius is approx. 1\%. Our mass limit is consistent with this hypothesis.

\textbf{TOI-745} was observed with HARPS, with 26 data points obtained showing no variability in phase with the transit ephemeris. Our joint fit gives a 95\% confidence upper limit on $K_b$ of 1.68 $\mathrm{m\,s}^{-1}$, corresponding to an upper limit on planetary mass of $2.48M_\oplus$. The RV data and model are shown in Figure \ref{fig:rvnulls}. The candidate radius is found to be $2.56\pm0.07R_\oplus$ under the planet model. No stellar companions were detected by high resolution imaging, although there are multiple nearby resolved sources in Gaia. The transit model shows some discrepancy in transit duration, implying that either a non-zero eccentricity would be required to better fit the data, or that our spectroscopic stellar density, which is included in the fit, is incorrect, perhaps because the transiting object orbits a different star with different parameters. Given the short orbital period of 1.08d, where the orbit would be expected to circularise rapidly, this casts further doubt on the planet hypothesis.

\begin{table*}
    \centering
    \caption{Fit results per system for new targets in this paper.}
    \begin{tabular}{llllllr}
        \hline
        \hline
        Star & Limb darkening & \textit{HARPS} offset ($\mathrm{m\,s}^{-1}$)& \textit{HARPS} jitter ($\mathrm{m\,s}^{-1}$)& \textit{PFS} offset ($\mathrm{m\,s}^{-1}$)& \textit{PFS} jitter ($\mathrm{m\,s}^{-1}$) & Linear drift($\mathrm{m\,s}^{-1}\mathrm{d}^{-1}$) \\
        \hline
       TOI-271 & [$0.92^{+0.06}_{-0.10}$ , $0.81^{+0.14}_{-0.33}$] & $61643.21^{+0.67}_{-0.67}$ & $0.40^{+0.74}_{-0.58}$ & $0.30^{+0.38}_{-0.39}$ & $2.87^{+0.37}_{-0.35}$ & -- \\
       TOI-510 & [$0.69^{+0.22}_{-0.33}$ , $0.51^{+0.32}_{-0.34}$] & $34926.47^{+0.63}_{-0.67}$ & $0.19^{+0.38}_{-0.15}$ &-- & -- & $0.785^{+0.070}_{-0.065}$ \\
       TOI-641 & [$0.85^{+0.11}_{-0.18}$ , $0.78^{+0.17}_{-0.34}$] & $37743.57^{+0.56}_{-0.57}$ & $0.60^{+0.94}_{-0.52}$ & $-0.42^{+0.53}_{-0.48}$ & $1.01^{+0.70}_{-0.58}$  & -- \\
       TOI-697 & [$0.39^{+0.36}_{-0.26}$ , $0.26^{+0.34}_{-0.18}$] & $26340.84^{+2.39}_{-2.41}$ & $5.62^{+2.52}_{-1.54}$ & --&--   & --\\
       TOI-745 & [$0.27^{+0.20}_{-0.13}$ , $0.63^{+0.26}_{-0.32}$] & $16915.34^{+0.62}_{-0.62}$ & $2.33^{+0.68}_{-0.62}$ & --&--  & --\\
        \hline
    \end{tabular}
    \label{tab:instparams}
\end{table*}

\begin{table*}
    \centering
    \caption{Fit results per candidate for new targets in this paper. All limits are 95\% confidence limits. $\rho_p$ limits calculated using $2\sigma$ lower radius value and the given mass limit. Radius and hence density values assume the candidate transits the expected host, which is likely not true in all cases. Further fit parameters are shown in Table \ref{tab:fit_app}.}
    \begin{tabular}{llllllllllr}
        \hline
        \hline
        Star & Candidate & Period (d) & Epoch (BJD-2458000) & K ($\mathrm{m\,s}^{-1}$) & $R_p/R_*$  & $R_p$ ($R_\oplus$) & $M_p$ ($M_\oplus$) &$\rho_p$ ($\mathrm{g\,cm}^{-3}$) \\
        \hline
       TOI-271 & b & $2.4759822\pm{0.0000040}$ & $382.22950^{+0.00091}_{-0.00092}$ & $<1.02$ & $0.0301^{+0.0032}_{-0.0038}$  & $4.16^{+0.50}_{-0.52}$ & $<2.39$ & $<0.43$ \\
       TOI-510 & b & $1.3523934\pm{0.0000010}$ & $469.58818^{+0.00068}_{-0.00070}$ & $0.66^{+0.35}_{-0.33}$ & $0.0304^{+0.0147}_{-0.0069}$  & $4.18^{+2.03}_{-0.96}$ & $1.08^{+0.58}_{-0.55}$ & -- \\
       TOI-510 & c & $10.02^{+1.33}_{-0.98}$ & $1168.78^{+1.48}_{-1.37}$ & $1.56^{+0.40}_{-0.41}$ & --  & -- & $4.82^{+1.29}_{-1.26}$ & --\\
       TOI-641 & b & $1.8930656\pm{0.0000049}$ & $1202.8169^{+0.0015}_{-0.0014}$ & $<0.77$ & $0.0316^{+0.0017}_{-0.0015}$  & $3.44^{+0.24}_{-0.22}$ & $<1.49$ & $<0.30$ \\
       TOI-697 & b &  $8.607856^{+0.000038}_{-0.000042}$ & $418.9270^{+0.0026}_{-0.0036}$ & $<9.60$ & $0.02145^{+0.00077}_{-0.00074}$  & $2.17^{+0.09}_{-0.08}$ & $<30.1$ & $<20.3$ \\
       TOI-745 & b & $1.0791049^{+0.0000031}_{-0.0000021}$ & $1330.9224^{+0.0009}_{-0.0014}$ & $<1.68$ & $0.02249^{+0.00059}_{-0.00061}$  & $2.56^{+0.07}_{-0.07}$ & $<2.48$ & $<0.97$\\
        \hline
    \end{tabular}
    \label{tab:fitresults}
\end{table*}

\section{Discussion} \label{sec:disc}
\begin{figure*}
    \centering
    \includegraphics[width=2\columnwidth]{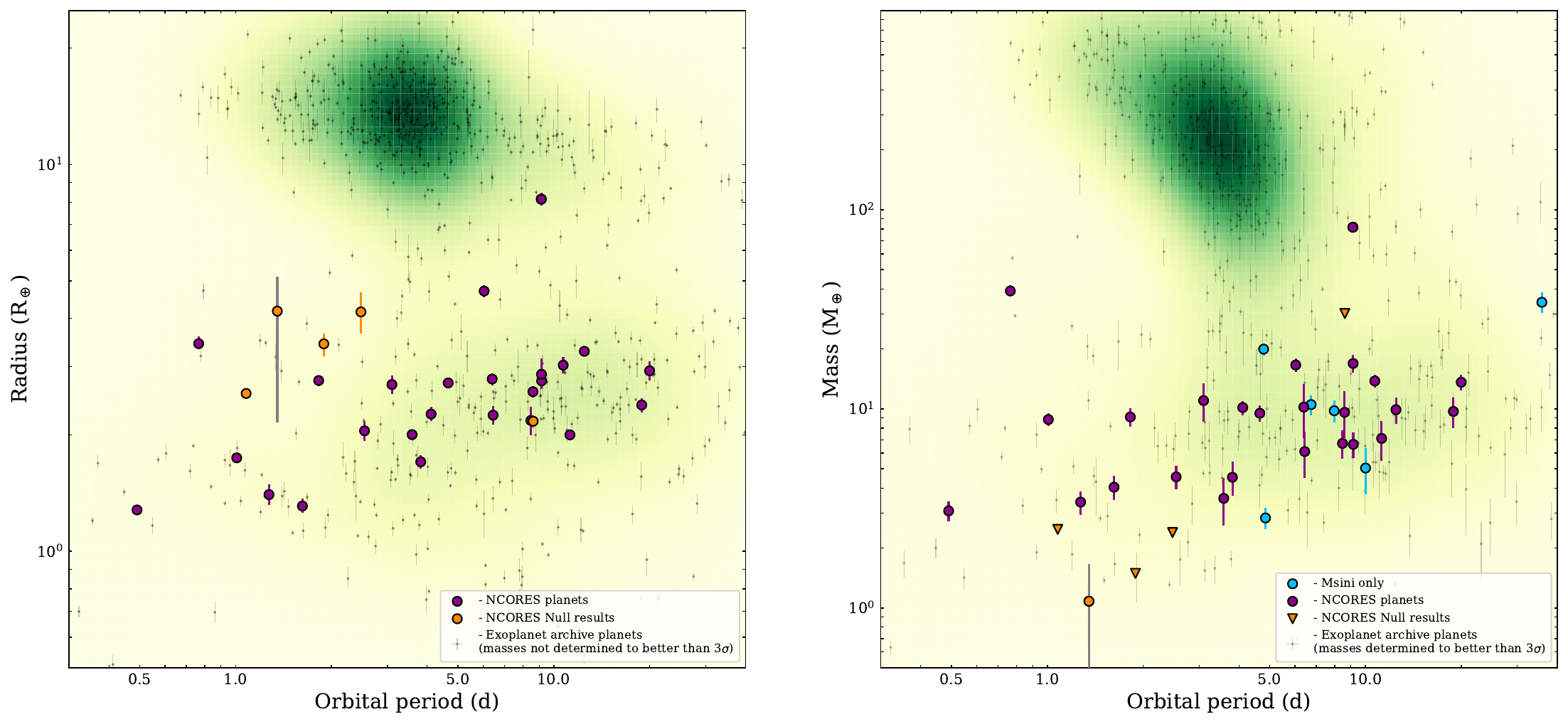}
    \caption{NCORES observed planets in orbital period - planet radius (left) and orbital period - planet mass (right). Confirmed planets are shown in purple and null results in orange. Planets detected in the radial velocities only and plotted with their $M_p\sin{i}$ values and shown in blue. A Gaussian kernel density estimate showing the background population of planets from the NASA Exoplanet Archive is shown in green.}
    \label{fig:NCORES_PRPM}
\end{figure*}

\begin{figure}
    \centering
    \includegraphics[width=0.82\columnwidth, trim={1.0cm 0.8cm 1.8cm 0.1cm}]{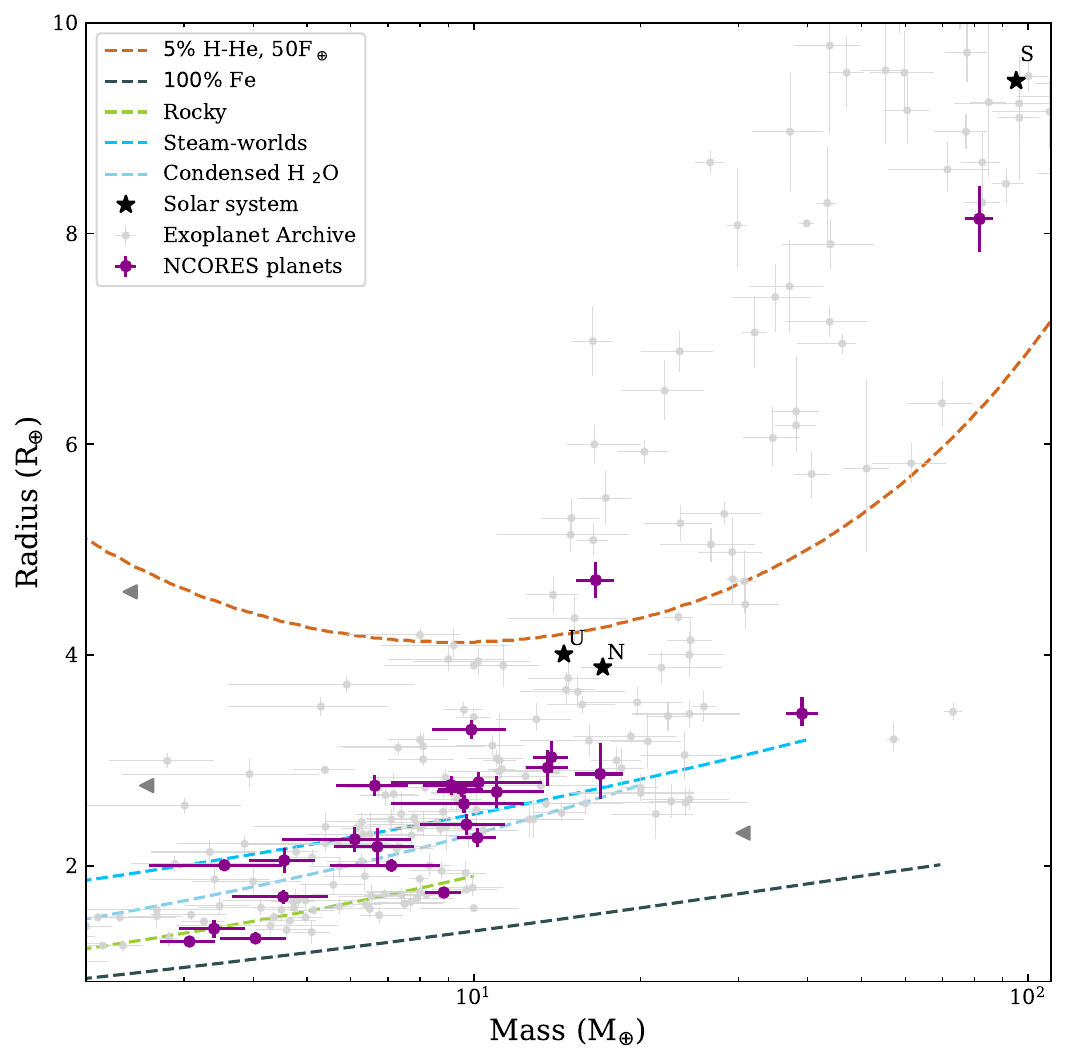}
    \caption{NCORES observed planets in mass-radius. Planets from the NASA Exoplanet Archive are shown for context in grey. Composition lines for pure iron \citep[dark grey,][]{Dorn15}, Earth-like (green), Steam-world (blue) and Condensed water-worlds (light blue) \citep{Venturini:2024} are shown, plus an additional line representing a planet with a 95 per cent water and a 5 per cent H-He envelope with $S=50S_{\oplus}$.}
    \label{fig:NCORE_MR}
\end{figure}

Having covered the less straightforward cases, here we move on to a broader interpretation of the whole NCORES sample, focused on confirmed planets published in our companion papers. Figure \ref{fig:NCORES_PRPM} shows the complete NCORES sample, in orbital period, planet radius and planet mass. The characterised planets span in size from Earths to sub-Saturns, and cover a broad range of compositions. The planet locations on the mass-radius diagram are shown in Figure \ref{fig:NCORE_MR}.

\subsection{Individual Highlights}
Highlights of the NCORES programme include the TOI-125 system, containing three super-Earth transiting planets near the 2:1 orbital resonance. The planets have similar size, $\sim 2.8R_\oplus$, but varying density and hence composition \citep{Nielsen:jq}, providing an interesting laboratory of planet formation. The TOI-431 system \citep{Osborn:fu} contains three detected planets, two transiting, with the inner, smaller planet on one side of the radius valley and an outer planet on the other, providing a chance to test theories of the valley's origin without needing to consider different host stars \citep{King:2024a}. Another key highlight is the TOI-849 system \citep{Armstrong:2020dm}, containing an ultrashort period Neptune-sized planet with a high mass, and density similar to the Earth. TOI-849b was the first `super-dense' Neptune discovered, and is hypothesised to be the remnant core of a gas giant planet, exposed through collisions, photoevaporation, or an unusual formation process which avoided runaway gas accretion. TOI-849b was one of the first planets discovered populating the Neptunian Desert.

\subsection{Planet core mass and composition distributions}
\label{sect:obsdists}
Studying the masses of planets near the radius gap, which are expected to have experienced substantial photoevaporation, was one of the goals of the survey. Theoretical models of the gap use the distribution of planet core masses and compositions as an input to the models, and also constrain those distributions based on the depth and location of the gap \citep[e.g][]{Rogers:2021ab}. This means that local exoplanet mass demographics are a key input to understanding the relevant physical processes. We supplement our NCORES sample by taking all confirmed planets from the NASA Exoplanet archive as of 11th January 2024 with mass and radius determinations better than $3\sigma$, and with masses and radii greater than 0.5 $M_\oplus$ and $R_\oplus$, respectively. We take the location of the gap in Instellation ($S$)-Radius($R_p)$ space, using the observationally derived relation from \citet{Ho:2023ab} of $\log (R_p/R_\oplus) = 0.07\log (S/S_\oplus) + 0.11$. Given this relation was determined for stars with $M_\star\ge0.7M_\odot$, we cut the sample to planets with host stars above this mass. Using instellation to define the gap allows the effect of varying luminosity from different host star stellar types on the evaporation potential of different planets to be reduced. If the radius gap is driven primarily by mechanisms other than photoevaporation, instellation may play a less important role, but will still combine with those mechanisms to some degree to produce the final planet sample.

\begin{figure}
    \centering
    \includegraphics[width=\columnwidth, trim={0 0.8cm 1.8cm 0.1cm}]{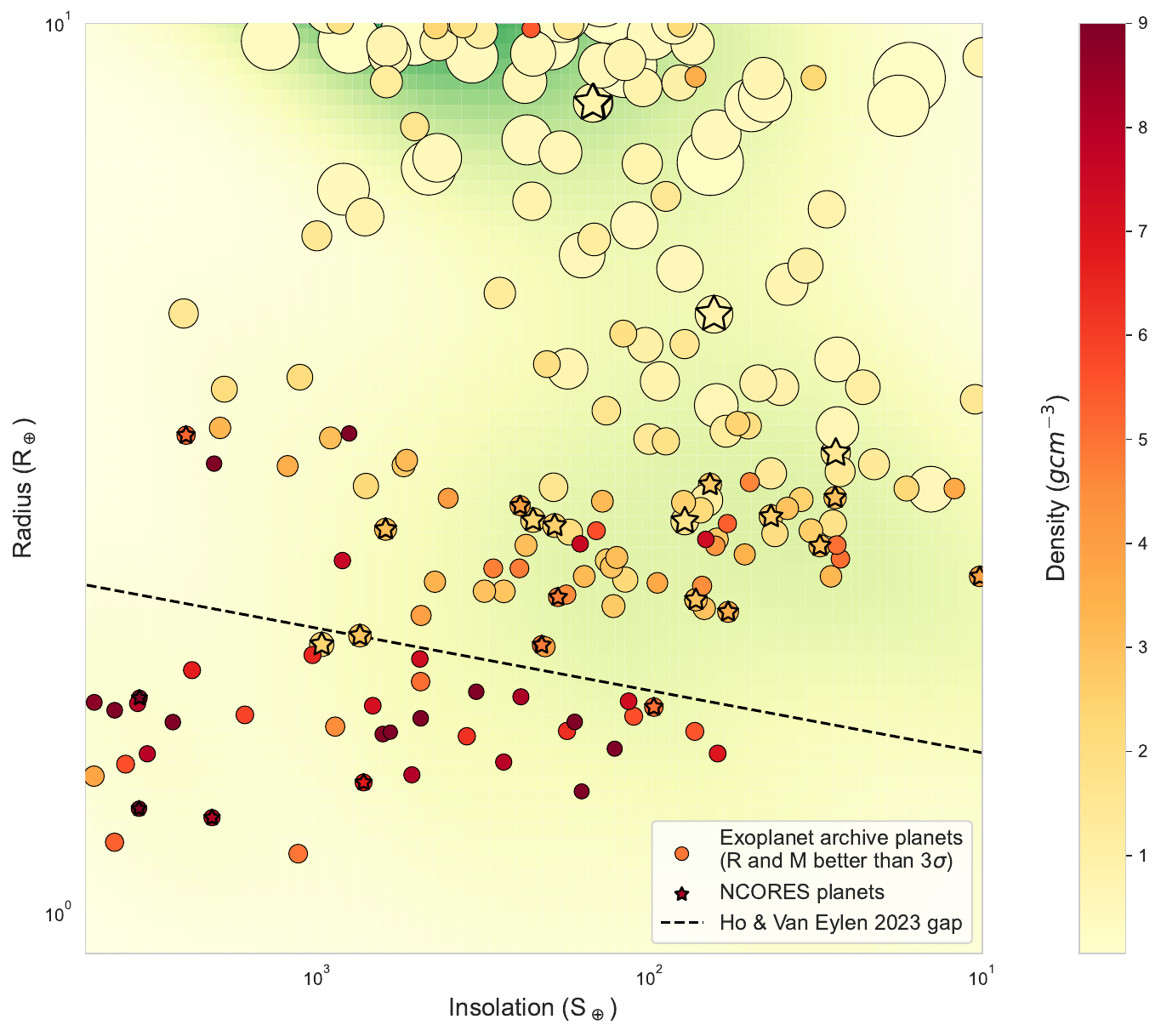}
    \caption{Planets with known mass and radius determined to better than $3\sigma$ plotted in Insolation-Radius space, coloured and inversely scaled in size by their density. Only planets with host star mass $M_*>0.7M_\odot$ are shown. The radius gap location determined in \citet{Ho:2023ab} is shown as a black dashed line. The background is coloured by a kernel density estimate of the planets shown. The diversity of sub-Neptune composition above the gap can be seen in the range of densities observed in this region.}
    \label{fig:SRplot}
\end{figure}

\begin{figure}
    \centering
    \includegraphics[width=0.8\columnwidth, trim={0 0.8cm 1.8cm 0.1cm}]{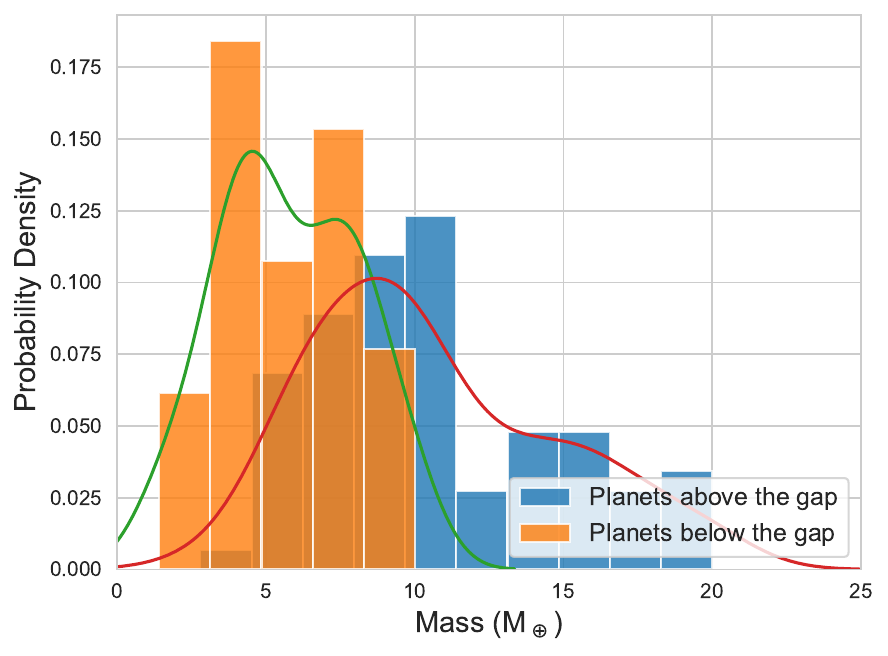}
    \caption{Distributions of planet masses below (orange) and above (blue) the radius gap drawn from Figure \ref{fig:SRplot}, with accompanying kernel density estimate curves. Only planets with mass $M_p<20M_\oplus$, host star mass $M_*>0.7M_\odot$ and Insolation $S_p>10S_\oplus$ are shown.}
    \label{fig:GapMassHists}
\end{figure}

Figure \ref{fig:SRplot} shows the retrieved sample and gap locations in Instellation-Radius space. Figure \ref{fig:GapMassHists} shows the distribution of planet masses below and above the gap, limited to planets with mass less than $20M_\oplus$ and instellation $S>10S_\oplus$ for clarity, giving a sample of 38 planets below the gap and 85 above for the FGK stars in our sample. The population of planets below the gap shows a distribution broadly consistent with the core mass distributions considered in \citet{Rogers:2021bc}, showing a broad peak of masses for planets below the gap (i.e. `bare cores') around $4M_\oplus$ that falls off at higher masses. Notably, the distribution also shows a sharp cutoff at $10M_\oplus$, sharper than seen in \citet{Rogers:2021bc}. 

\begin{figure}
    \centering
    \includegraphics[width=0.8\columnwidth, trim={0 0.8cm 1.8cm 0.1cm}]{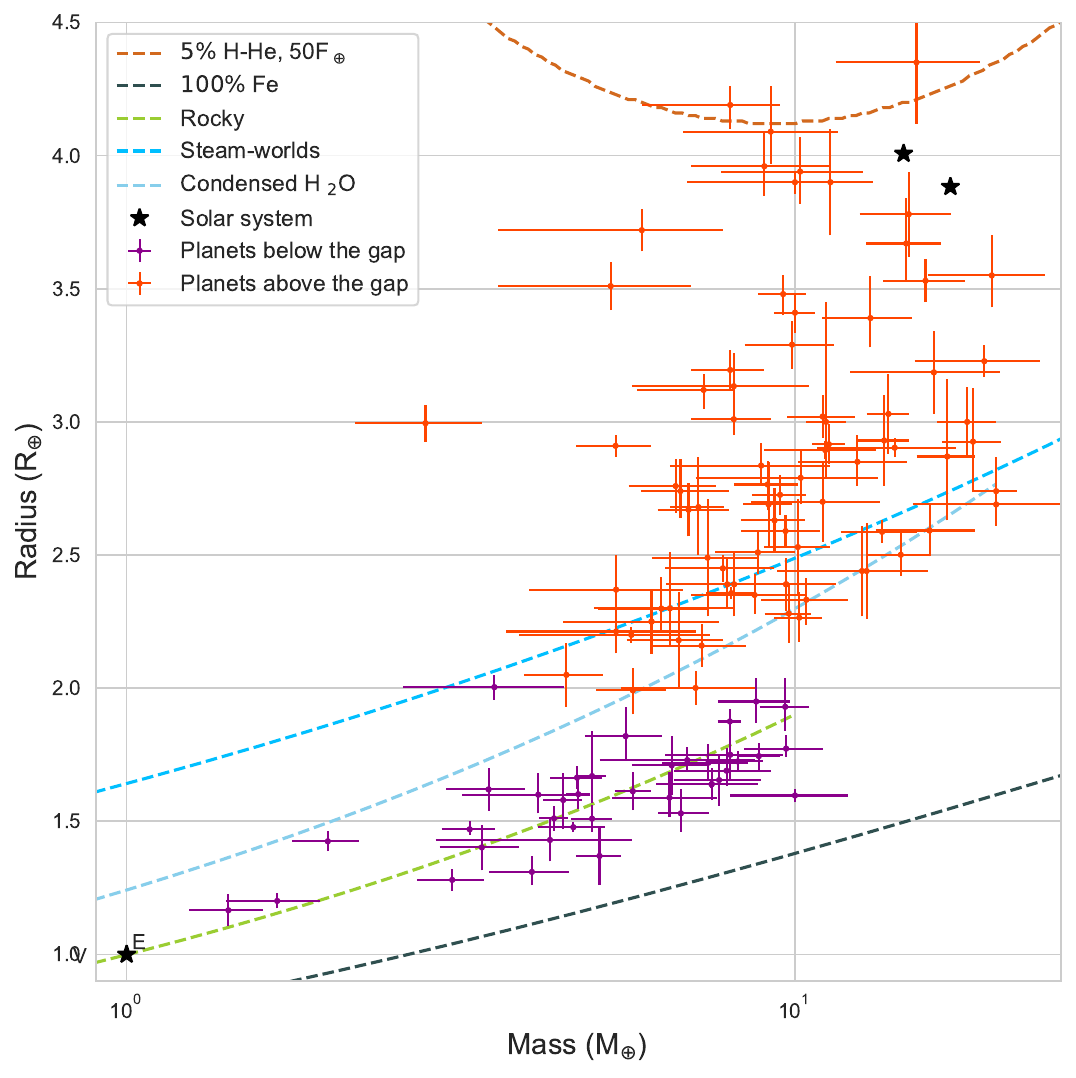}
    \caption{Planets below (purple) and above (orange) the radius gap drawn from Figure \ref{fig:SRplot}, in mass-radius space. The same composition lines as in Figure \ref{fig:NCORE_MR} are shown. Only planets with host star mass $M_*>0.7M_\odot$ are shown. The outlier point below the gap yet compatible with a 'steam-world' composition is TOI-396 b.}
    \label{fig:GapMR}
\end{figure}

Figure \ref{fig:GapMR} shows the same sample on the mass-radius diagram. Planets below the gap are strongly consistent with an Earth-like composition, in agreement with models of photoevaporation, core-powered mass loss, and differing formation channels. Planets above the gap show a much more divergent set of compositions, even within the 20$M_\oplus$ upper limit considered, implying a mixture of formation channels as suggested in \citet{Ho:2024ab}.

It is important to note that this sample contains many observational biases, particularly driven by the smaller radial velocity signal expected for lower mass and/or longer orbital period planets, as well as targeting and observation choices by different teams. As such, details of the distribution should be treated with caution. Nonetheless, the cutoff at $10M_\oplus$, the strong tendency for planets below the gap to match an Earth-like composition \citep[with the curious exception of TOI-396b,][]{2024arXiv241114911B}, and the diversity in composition of planets above the gap are all clear features which would not be expected to be removed by observational biases.

\subsection{Comparison to planetary formation and evolution models}
\label{sec:formation}
To give context to the observed distribution of planetary masses, we created two theoretical samples from simulations of planet formation and evolution. `Model 1' is based on planetesimal accretion simulations from \citet{Emsenhuber:2021a,Emsenhuber:2021b} and \citet{Burn:2021,Burn:2024}, whereas `Model 2' corresponds to pebble accretion simulations from \citet{Venturini:2024} \citep[based on the model from][]{Venturini:2020rvalley}. Both models have in common that they compute the growth of planets from a moon-mass stage by solid and gas accretion, including the evolution of the gaseous disc and considering as well orbital migration. The evolution phase starts at the time of disc dispersal and  calculates the cooling of the planets in a Gyr-timescale including atmospheric-mass loss by photoevaporation. The planets that end up below the radius gap are bare rocky cores whose atmospheres were stripped by photoevaporation, while planets above the gap formed beyond the ice line and could retain an envelope of (supercritical) vapour mixed with some (or none) hydrogen and helium \citep{Venturini:2020rvalley, Burn:2024}. 
For Model 1, the solids accreted correspond to 300 meter planetesimals, while for Model 2, to pebbles evolving from micro-meter dust to cm-grains. The models have several other differences, namely Model 1 considers 100 gravitating, moon-mass embryos per disc as an initial condition, while Model 2 follows the growth of one 1 embryo per disc. More details about the models can be found in the corresponding references.

The synthetic planets are separated into planets above and below the gap using the same criteria as the observed sample. 
By randomly drawing synthetic planets weighted by the observed stellar mass distribution, transit probability, and an estimate of the radial velocity semi-amplitude required, the synthetic planets' mass distribution can be compared to the observed sample discussed in Section \ref{sect:obsdists}.

\begin{figure}
    \centering
    \includegraphics[width=\linewidth]{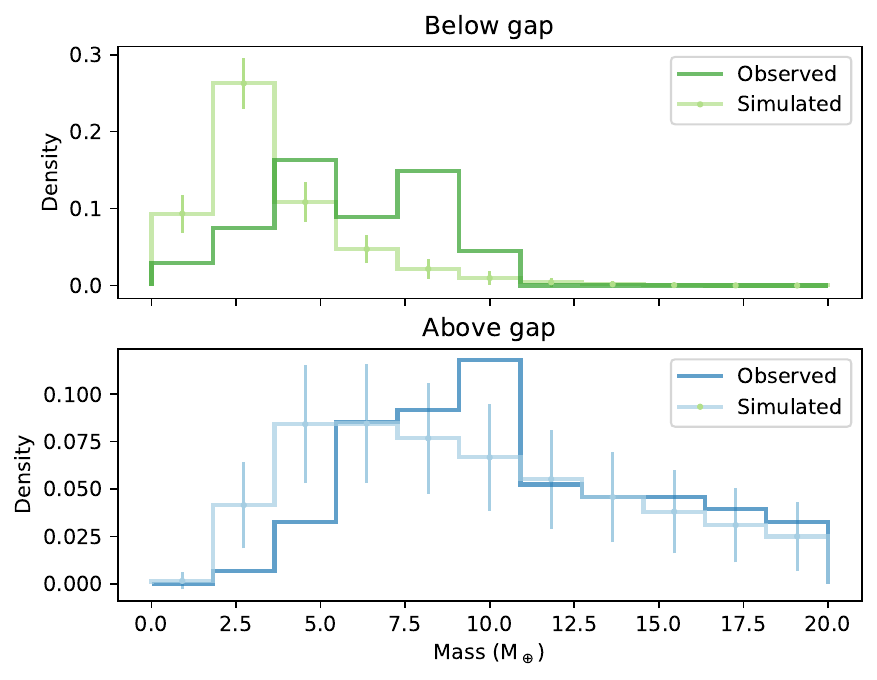}
    \caption{Mass distribution comparison of observed and synthetic planets (planetesimals-based models). Planets are categorised into groups above and below the valley as described in the text. The simulated distribution of planets is from an extension of the planet evolution model presented in \citet{Burn:2024} to synthetic populations around lower stellar masses \citep{Burn:2021}. For the comparison, an error is randomly applied to the theoretical planetary masses, a minimum mass of 1.4 $M_\oplus$ and period of 30 days are required, and several samples of 121 planets are drawn using the synthetic planets transit \citep[following][]{Petigura:2018CalKep4} and radial velocity (logistic function centered at semi-amplitude of 5 $\mathrm{m\,s}^{-1}$ with growth rate of 2 $\mathrm{m\,s}^{-1}$) detection probability. The errorbars show the resulting standard deviation of the bin value over 5000 draws.}
    \label{fig:mass_model-1_comparison}
\end{figure}
\begin{figure}
    \centering
    \includegraphics[width=\linewidth]{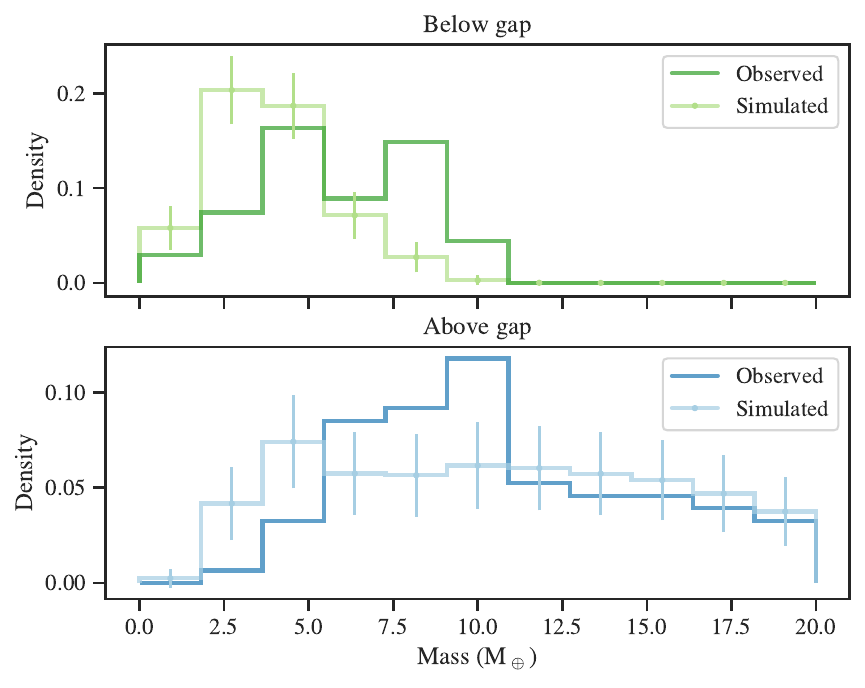}
    \caption{Mass distribution comparison of observed and synthetic planets (pebbles-based models). Planets are categorised into groups above and below the valley as described in the text. The simulated distribution of planets stems from the planet formation and evolution models of \citet{Venturini:2024} for stellar masses of 0.7 and 1.0 \msun. For the comparison, we consider planets in the mass range 0.5< M$_{\rm P}$ < 20 $M_\oplus$ and insolation S>10 S$_{\oplus}$. An error is randomly applied to the theoretical planetary masses, and several samples of 121 planets are drawn using the synthetic planets transit \citep[following][]{Petigura:2018CalKep4} and radial velocity (logistic function centered at semi-amplitude of 5 $\mathrm{m\,s}^{-1}$ with growth rate of 2 $\mathrm{m\,s}^{-1}$) detection probability. The errorbars show the resulting standard deviation of the bin value over 5000 draws.}
    \label{fig:mass_model-2_comparison}
\end{figure}

The synthetic distributions in Figures \ref{fig:mass_model-1_comparison} and \ref{fig:mass_model-2_comparison} show the same qualitative trend to observations: planets below the gap have typically lower masses than planets above the gap. 
While a good general agreement is found between observations and the models, regarding the planets below the gap, both models predict a peak of super-Earths at lower masses compared to observations, with the peak of Model 2 closer to the observed one. The excess of the lowest-mass planets could be explained by the difficulty to detect such planets. If more low-mass planets were detected in the future, the second peak of the observed distribution at $\sim$8 M$_\oplus$ could shift to lower values. This would also alleviate the current problem of the deficit of synthetic planets at $\sim$8 M$_\oplus$. Alternatively, the deficit of synthetic super-Earths at $\sim$8 M$_\oplus$ could indicate some missing process or some inadequate choice of physical parameter in the formation-evolution modelling. For example, rocky cores with masses above 5 M$_\oplus$ can in principle be produced by pebble accretion for discs with extremely low viscosity \citep[$\alpha$-viscosity < $10^{-4}$, see][]{Venturini:2020pebblemasslimit}, a value not typically adopted in standard core accretion simulations. Those massive cores have the additional problem of being prone to accrete gas quickly, rendering the planets as giants. Processes like atmospheric recycling during the gas-disc phase \citep[e.g.][]{Ormel:15, Moldenhauer:23} or giant impacts after the disc dispersal \citep[e.g.][]{Inamdar15} might contribute to keep massive super-Earths without atmospheres for M$_{\rm P}<10$ M$_\oplus$. Alternatively, and considering that these massive super-Earths are rather hot \citep[equilibrium temperatures larger than 1000 K, see ][]{Parc:24}, these objects could be produced by the photoevaporation of steam atmospheres, a process that is at the moment poorly modelled and which deserves further investigation. 

Regarding the masses of the sub-Neptunes (bottom panels of Figs.\ref{fig:mass_model-1_comparison} and \ref{fig:mass_model-2_comparison}), both models show a deficit of planets at the observed peak location of $\sim$10 M$_\oplus$. This could be related to the problem of rapid gas accretion at this mass range that we stated above. 
In addition, Model 1 has the peak of masses at lower values than the observed planets, while for Model 2 the distribution of sub-Neptunes is more spread, with a median at $\sim$10.4 M$_{\oplus}$ (slightly larger than the observed one at $\sim$9.8 M$_{\oplus}$). The higher median masses of Model 2 relative to Model 1 (at $\sim$8.8 M$_{\oplus}$) are probably reflecting the enhanced pebble accretion rate beyond the ice line compared to planetesimal accretion. This might indicate an interesting direction to explore to distinguish between planetesimal and pebble accretion models once a larger sample of planet masses is measured.


Indeed, at the moment it is not possible to distinguishing between planetesimals- vs. pebbles-based models, the current mass distributions are too similar for observations to prefer one over the other. To make progress on the observational side, it is key to increase the sample of planets with precise masses. On the theoretical side, it is important to perform more sensitivity studies to test broader ranges of model parameters, including as well new physical processes such as atmospheric recycling or a longer phase of the giant impact stage \citep[see, e.g.][]{Emsenhuber:2021a}. It is also important to set-up equal initial condition studies \citep[e.g.][]{Brugger20}, to properly disentangle if the differences in the obtained distributions stem from the solid accretion (planetesimals vs. pebbles) or other model set-ups (e.g. number and injection time of planetary seeds, envelope opacities, which are also different between the two models employed here).     

Overall, planets below and above the gap are starting to exhibit distinct mass distributions, with sub-Neptunes being on average more massive than super-Earths. This can be understood by formation models due to the presence of the ice line and due to the migration process. The ice line enhances the surface density of solids, and increases the pebble accretion efficiency \citep{Venturini:2020rvalley}, producing larger icy than rocky cores (both for planetesimals and pebbles, but the effect is stronger for pebble accretion). Planets formed beyond the ice line not only tend to be more massive, but also, only the planets exceeding a certain mass threshold can undergo type-I migration and reach close-in orbits \citep{Paardekooper2011, Burn:2021}. Thus, migration enhances the separation of lighter rocky super-Earths versus more massive, water-rich sub-Neptunes.
On the other hand, pure evolution models postulate that super-Earths and sub-Neptunes were the same kind of planets after formation, with the same underlying core mass distribution and composition \citep[e.g.][]{Owen:2017kf, Gupta:2019ab, Rogers:2021bc}. A prediction from those models is that sub-Neptunes only differentiate from super-Earths regarding the extra 1\% by mass of H/He that could be retained during the atmospheric erosion \citep{Owen:2017kf}. In principle, this would have a negligible impact in the mass distributions of the two planet types. However, atmospheric escape depends on the planet mass, implying that a mass separation due to atmospheric escape alone could also be expected. While atmospheric escape simulations show a general agreement in mass-radius with the observed population \citep{Kubyshkina22}, a significant fraction of sub-Neptunes cannot be reproduced by atmospheric escape alone under the assumption of a unique core mass distribution \citep{Kubyshkina22}.

\subsection{Biases and Future Work}
\label{sect:biases}
While this programme provided a large number of useful individual systems to test planetary formation theories, and increased the sample of precisely characterised planets in this area of parameter space, the next step in building an observational picture of the exoplanet population is to generate a statistically robust sample. The sample presented in Figure \ref{fig:NCORES_PRPM} was compiled by selecting bright, well-vetted candidates on a rolling basis as they were detected by the \textit{TESS} mission. There are many underlying biases, including human bias from the vetting and selection methods, number and availability of follow-up observations, degree of follow-up before a system was rejected, and general observational biases towards short orbital periods and larger planets due to their increased radial velocity amplitude.

In addition, the desire to reach a set confidence level, often $3\sigma$ or $5\sigma$, before publishing is known to lead to a potential bias in the distribution of small planet masses \citep{Montet2018aa,Burt2018bb}. The best way to avoid such bias is to publish low-significance and null results, allowing mass limits to be incorporated in statistical population studies \citep{Wolfgang2016aa}.

Overcoming these biases and limitations of typical RV surveys to produce a sample useful for direct comparison to formation and evolution models is an important goal for the field. We deliberately publish our null results with the aim of helping move towards this goal in the longer term, and enable future sample-level studies. Adoption of a merit function and quantified target selection, as seen in \citet{Teske:2021ab}, is a strong further step which we have introduced in future surveys.



\section{Conclusions}
We have presented an overview of results from the NCORES observation programme characterising super-Earth and Neptune size planets with HARPS. In addition, in this paper, we:
\begin{itemize}
    \item Derive stellar parameters from high-resolution HARPS spectra for five candidate-hosting stars observed by \textit{TESS}.
    \item Publish radial velocities of those targets and upper limits on the mass of four \textit{TESS} candidate planets.
    \item Present a marginal mass measurement for the candidate TOI-510b, implying an Earth-like mass but significantly lower density, and the discovery of a new non-transiting candidate planet TOI-510c.
    \item Show the current distribution of planet masses and radii around the radius gap, and compares the distribution to a combined formation and evolution model.

With this work we hope to encourage the publication of null results where possible from large exoplanet surveys, helping to enable more complete statistical studies in future. 
\end{itemize}

\section*{Data Availability Statement}
HARPS and PFS radial velocities are published in Tables \ref{tab:spec1} and \ref{tab:specpfs}, respectively. Raw spectra are available from the ESO public archive\footnote{http://archive.eso.org/}. All follow-up data are available on the EXOFOP-TESS website\footnote{https://exofop.ipac.caltech.edu/tess}. All \textit{TESS} lightcurve data are available from the Michulski Archive for Space Telescopes\footnote{https://mast.stsci.edu/}.

\section*{Acknowledgements} \footnotesize{
The authors would like to thank the referee for helpful comments which improved the paper. Based on observations collected at the European Organisation for Astronomical Research in the Southern Hemisphere under ESO programme 1102.C-0249.
This paper includes data collected by the \textit{TESS} mission. Funding for the \textit{TESS} mission is provided by the NASA Explorer Program. Resources supporting this work were provided by the NASA High-End Computing (HEC) Program through the NASA Advanced Supercomputing (NAS) Division at Ames Research Center for the production of the SPOC data products. We acknowledge the use of public \textit{TESS} Alert data from pipelines at the \textit{TESS} Science Office and at the \textit{TESS} Science Processing Operations Center.
This work makes use of observations from the LCOGT network. Part of the LCOGT telescope time was granted by NOIRLab through the Mid-Scale Innovations Program (MSIP). MSIP is funded by NSF.
This paper makes use of observations made with the MuSCAT2 instrument, developed by the Astrobiology Center, at TCS operated on the island of Tenerife by the IAC in the Spanish Observatorio del Teide.
This paper makes use of data from the MEarth Project, which is a collaboration between Harvard University and the Smithsonian Astrophysical Observatory. The MEarth Project acknowledges funding from the David and Lucile Packard Fellowship for Science and Engineering, the National Science Foundation under grants AST-0807690, AST-1109468, AST-1616624 and AST-1004488 (Alan T. Waterman Award), the National Aeronautics and Space Administration under Grant No. 80NSSC18K0476 issued through the XRP Program, and the John Templeton Foundation.
This research has made use of the Exoplanet Follow-up Observation Program (ExoFOP; DOI: 10.26134/ExoFOP5) website, which is operated by the California Institute of Technology, under contract with the National Aeronautics and Space Administration under the Exoplanet Exploration Program.
Some of the observations in this paper made use of the High-Resolution Imaging instruments ‘Alopeke and Zorro obtained under Gemini LLP Proposal Number: GN/S-2021A-LP-105. ‘Alopeke and Zorro were funded by the NASA Exoplanet Exploration Program and built at the NASA Ames Research Center by Steve B. Howell, Nic Scott, Elliott P. Horch, and Emmett Quigley. They are mounted on the Gemini North and South telescopes respectively of the international Gemini Observatory, a program of NSF’s OIR Lab, which is managed by the Association of Universities for Research in Astronomy (AURA) under a cooperative agreement with the National Science Foundation. on behalf of the Gemini partnership: the National Science Foundation (United States), National Research Council (Canada), Agencia Nacional de Investigación y Desarrollo (Chile), Ministerio de Ciencia, Tecnología e Innovación (Argentina), Ministério da Ciência, Tecnologia, Inovações e Comunicações (Brazil), and Korea Astronomy and Space Science Institute (Republic of Korea).




This research was funded in part by the UKRI, (Grants ST/X001121/1, EP/X027562/1). R.B. acknowledges the support from the German Research Foundation (DFG) under Germany’s Excellence Strategy EXC 2181/1-390900948, Exploratory project EP 8.4 (the Heidelberg STRUCTURES Excellence Cluster). Part of this research was carried out at the Jet Propulsion Laboratory, California Institute of Technology, under a contract with the National Aeronautics and Space Administration (80NM0018D0004). Co-funded by the European Union (ERC, FIERCE, 101052347). Views and opinions expressed are however those of the author(s) only and do not necessarily reflect those of the European Union or the European Research Council. Neither the European Union nor the granting authority can be held responsible for them. This work was supported by FCT - Fundação para a Ciência e a Tecnologia through national funds and by FEDER through COMPETE2020 - Programa Operacional Competitividade e Internacionalização by these grants: UIDB/04434/2020; UIDP/04434/2020. We acknowledge financial support from the Agencia Estatal de Investigaci\'on of the Ministerio de Ciencia e Innovaci\'on MCIN/AEI/10.13039/501100011033 and the ERDF “A way of making Europe” through project PID2021-125627OB-C32, and from the Centre of Excellence “Severo Ochoa” award to the Instituto de Astrofisica de Canarias. The work of HPO has been carried out within the framework of the NCCR PlanetS supported by the Swiss National Science Foundation under grants 51NF40\_182901 and 51NF40\_205606. DB and JLB are funded by the MICIU/AEI/10.13039/501100011033  PID2019-107061GB-C61 grant and NextGenerationEU/PRTR grant CNS2023-144309. X.D acknowledges the support from the European Research Council (ERC) under the European Union’s Horizon 2020 research and innovation programme (grant agreement SCORE No 851555) and from the Swiss National Science Foundation under the grant SPECTRE (No 200021\_215200). We thank the Swiss National Science Foundation (SNSF) and the Geneva University for their continuous support. This work has been carried out within the framework of the NCCR PlanetS supported by the Swiss National Science Foundation under grants 51NF40\_182901 and 51NF40\_205606. PJW acknowledges support from the UK Science and Technology Facilities Council (STFC) through consolidated grants ST/T000406/1 and ST/X001121/1. JV acknowledges support from the Swiss National Science Foundation (SNSF) under grant PZ00P2\_208945. C.M. acknowledges the support from the Swiss National Science Foundation under grant 200021\_204847 `PlanetsInTime'. E.D.M. acknowledges support by the Spanish MICIN/AEI/10.13039/50110001103 and the Ramón y Cajal fellowship RyC2022-035854-I. D.B., J.L.B., E.D.M. are funded by the MICIN/AEI/PID2023-150468NB-I00.
}

\section*{Affiliations}
\footnotesize{
$^{1}$Department of Physics, University of Warwick, Coventry CV4 7AL, UK\\
$^{2}$Centre for Exoplanets and Habitability, University of Warwick, Gibbet Hill Road, Coventry CV4 7AL, UK\\
$^{3}$Department of Physics and Astronomy, McMaster University, 1280 Main St W, Hamilton, ON, L8S 4L8, Canada\\
$^{4}$Max Planck Institute for Astronomy, Königstuhl 17, 69117 Heidelberg, Germany\\
$^{5}$Department of Astronomy of the University of Geneva, 51 chemin de Pegasi, 1290 Versoix, Switzerland\\
$^{6}$Instituto de Astrof\'isica e Ci\^encias do Espa\c{c}o, Universidade do Porto, CAUP, Rua das Estrelas, 4150-762 Porto, Portugal\\
$^{7}$Austrian Academy of Sciences, Schmiedlstrasse 6, A-8042 Graz, Austria\\
$^{8}$ Jet Propulsion Laboratory, California Institute of Technology, 4800 Oak Grove Drive, Pasadena, CA 91109, USA\\
$^{9}$Center for Astrophysics \textbar \ Harvard \& Smithsonian, 60 Garden Street, Cambridge, MA 02138, USA\\
$^{10}$Centro de Astrobiolog\'ia (CAB, CSIC-INTA), Depto. de Astrof\'isica, ESAC campus, 28692, Villanueva de la Ca\~nada (Madrid), Spain\\
$^{11}$NASA Ames Research Center, Moffett Field, CA 94035, USA\\
$^{12}$George Mason University, 4400 University Drive, Fairfax, VA, 22030 USA\\
$^{13}$American Association of Variable Star Observers, 185 Alewife Brook Parkway, Suite 410, Cambridge, MA 02138, USA\\
$^{14}$Department of Physics and Astronomy, University of Kansas, Lawrence, KS, USA\\
$^{15}$International Center for Advanced Studies (ICAS) and ICIFI (CONICET), ECyT-UNSAM, Campus Miguelete, 25 de Mayo y Francia, (1650) Buenos Aires, Argentina\\
$^{16}$Tsung-Dao Lee Institute, Shanghai Jiao Tong University, 1 Lisuo Road, Shanghai 201210, People’s Republic of China\\ 
$^{17}$School of Physics and Astronomy, Shanghai Jiao Tong University, 800 Dongchuan Road, Shanghai 200240, People’s Republic of China\\
$^{18}$U.S. Naval Observatory, Washington, D.C. 20392, USA\\
$^{19}$Max-Planck-Institut f\"ur Astronomie, K\"onigstuhl 17, 69117 Heidelberg, Germany\\
$^{20}$Space Research and Planetary Sciences, Physics Institute, University of Bern, Gesellschaftsstrasse 6, 3012 Bern, Switzerland\\
$^{21}$Instituto de Astrof\'isica de Canarias (IAC), 38200 La Laguna, Tenerife, Spain\\
$^{22}$Deptartamento de Astrof\'isica, Universidad de La Laguna (ULL), 38206 La Laguna, Tenerife, Spain\\
$^{23}$Departamento de F\'isica e Astronomia, Faculdade de Ci\^encias, Universidade do Porto, Rua do Campo Alegre, Porto, Portugal\\
$^{24}$Department of Physics and Astronomy, Vanderbilt University, Nashville, TN 37235, USA\\
$^{25}$Perth Exoplanet Survey Telescope, Perth, Western Australia, Australia\\
$^{26}$Earth and Planets Laboratory, Carnegie Institution for Science \& The Observatories of the Carnegie Institution for Science, 5241 Broad Branch Rd NW, Washington, DC 20015 \& 813 Santa Barbara Street, Pasadena, CA, 91101\\
$^{27}$Department of Physics \& Astronomy, Johns Hopkins University, 3400 N. Charles Street, Baltimore, MD 21218, USA
}



\bibliographystyle{mnras}
\bibliography{papers061219}


\appendix
\section{Additional Figures and Tables}

\begin{figure}
    \centering
    \includegraphics[width=\columnwidth]{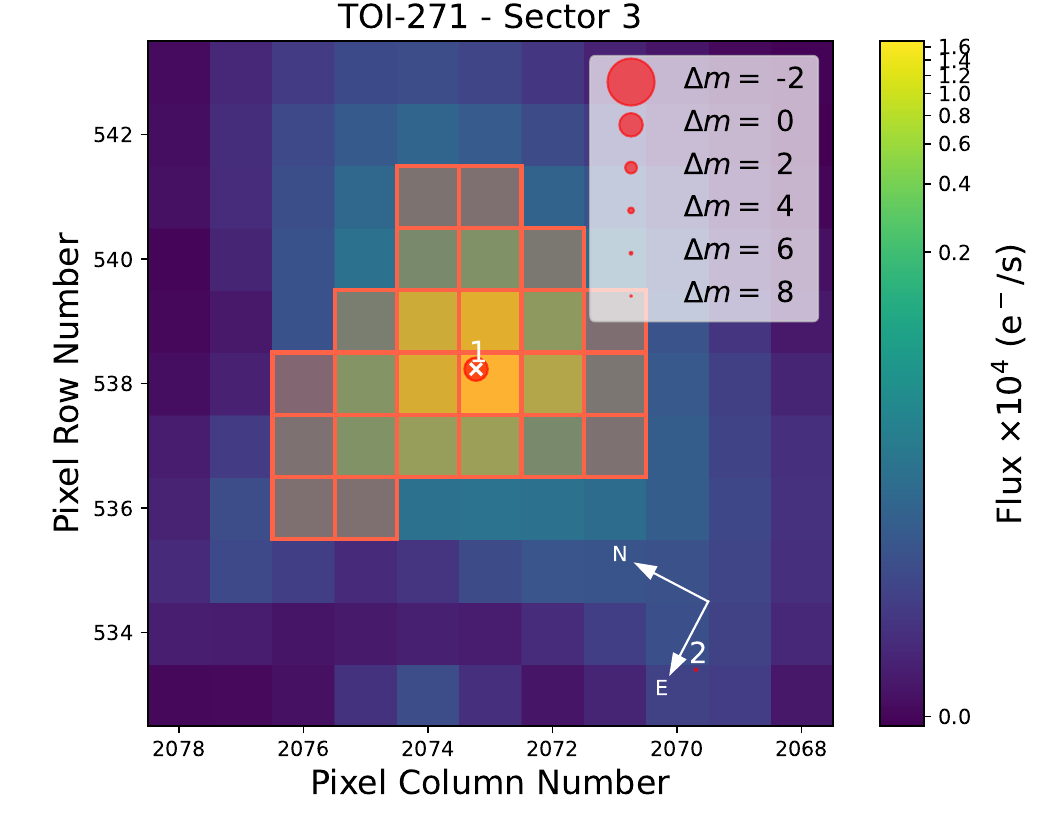}
    \caption{As Figure \ref{fig:tpf510} for TOI-641 in Sector 6.}
    \label{fig:tpf271}
\end{figure}

\begin{figure}
    \centering
    \includegraphics[width=\columnwidth]{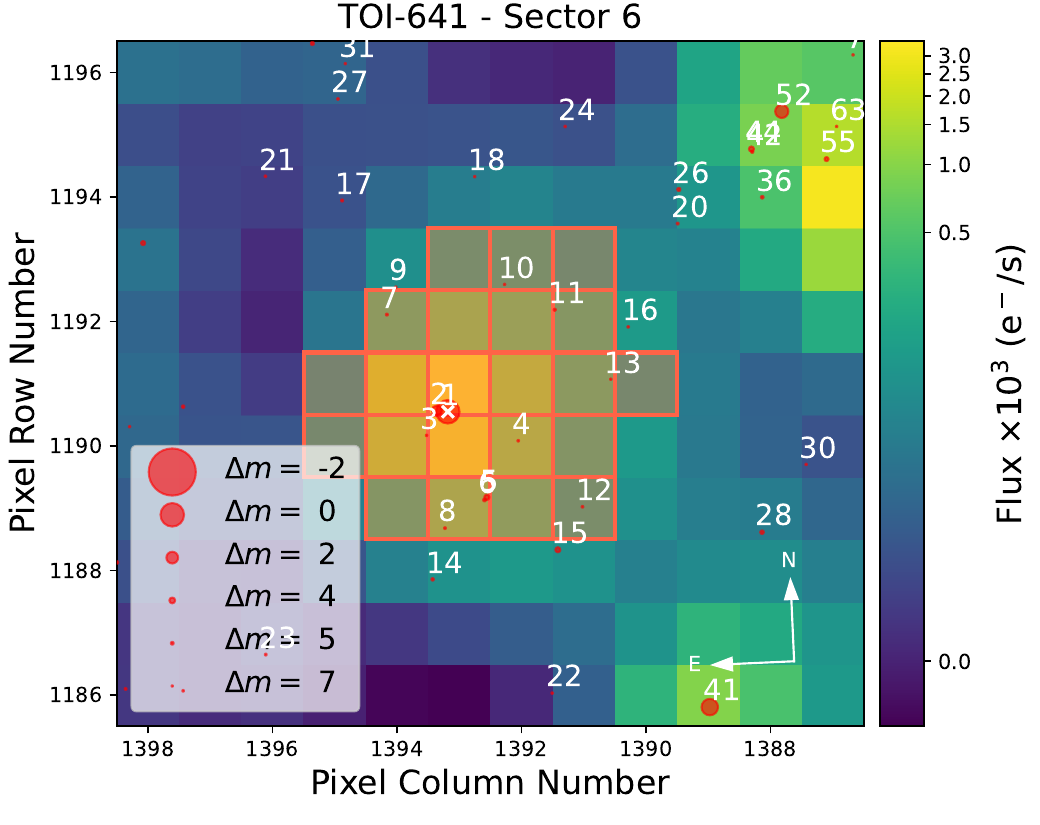}
    \caption{As Figure \ref{fig:tpf510} for TOI-641 in Sector 6.}
    \label{fig:tpf641}
\end{figure}

\begin{figure}
    \centering
    \includegraphics[width=\columnwidth]{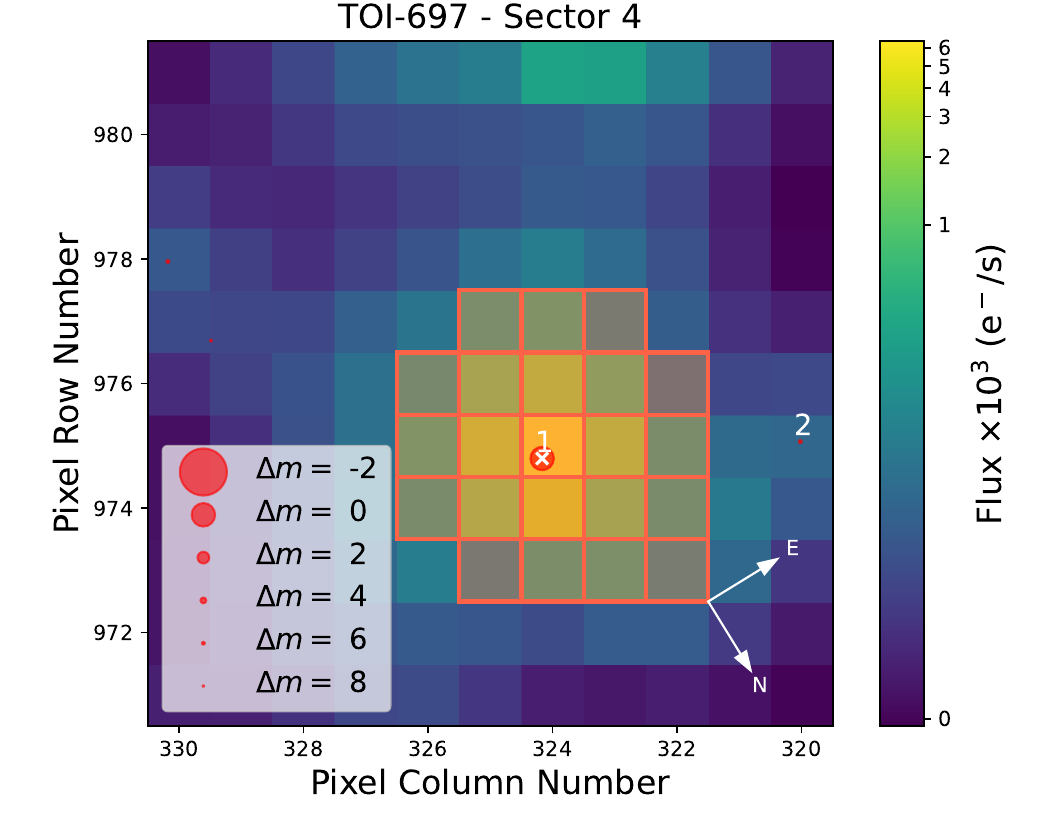}
    \caption{As Figure \ref{fig:tpf510} for TOI-697 in Sector 4.}
    \label{fig:tpf697}
\end{figure}

\begin{figure}
    \centering
    \includegraphics[width=\columnwidth]{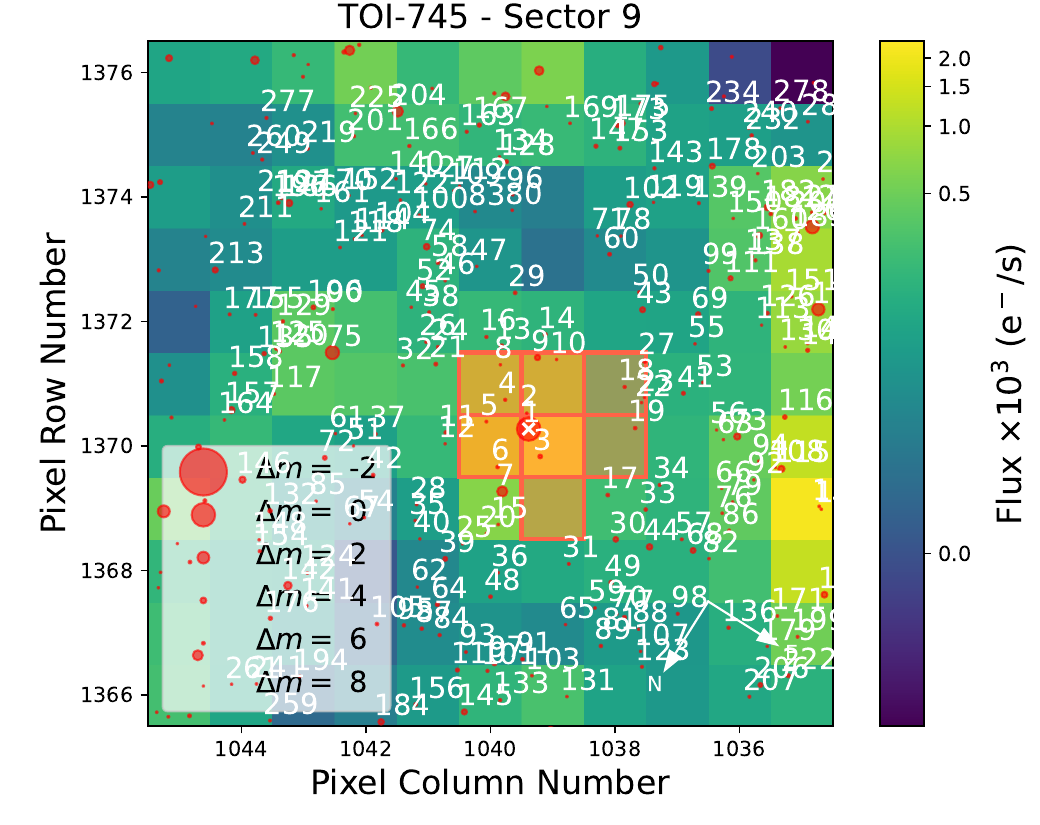}
    \caption{As Figure \ref{fig:tpf510} for TOI-745 in Sector 9.}
    \label{fig:tpf745}
\end{figure}


\begin{table*}
\caption{HARPS spectroscopy. Full table available online.}
\label{tab:spec1}
\begin{tabular}[c]{lrrrrrrrr}
\hline
\hline
Object & Time & RV & $\sigma_{\rm RV}$  & FWHM & contrast & bis span & $S_{\rm MW}$ & H$\alpha$ \\
& $[\rm{BJD}-2458000]$ & $\mathrm{m\,s}^{-1}$ & $\mathrm{m\,s}^{-1}$ & $\mathrm{m\,s}^{-1}$  & \multicolumn{4}{c}{} \\
\hline
TOI-271 & 474.664039 & 61644.563081 & 1.241512 & 8728.145 & 32.413 & 38.699 & 0.152 & 0.228\\
TOI-271 & 474.754847 & 61642.956125 & 1.413570 & 8719.052 & 32.418 & 36.577 & 0.151 & 0.226\\
TOI-271 & 475.685012 & 61640.967255 & 1.366993 & 8719.110 & 32.446 & 40.034 & 0.151 & 0.230\\
\vdots & \vdots & \vdots & \vdots & \vdots & \vdots & \vdots & \vdots & \vdots \\
\hline
\hline
\end{tabular}
\end{table*}


\begin{table}
\caption{PFS spectroscopy. Full table available online.}
\label{tab:specpfs}
\begin{tabular}{lrrr}
\hline
\hline
Object & Time & RV & $\sigma_{\rm RV}$  \\
& $[\rm{BJD}-2458000]$ & $\mathrm{m\,s}^{-1}$ & $\mathrm{m\,s}^{-1}$ \\
\hline
TOI-271 & 467.541910 & 1.67 & 2.71\\
TOI-271 & 467.549540 & -1.38 & 2.62\\
TOI-271 & 467.747260 & 1.50 & 2.17\\
\vdots & \vdots & \vdots & \vdots \\
\hline
\hline
\end{tabular}
\end{table}

\begin{table*}
    \centering
    \caption{Fit results not shown in Table \ref{tab:fitresults}.}
    \begin{tabular}{llllllr}
        \hline
        \hline
        Star & $b$ & Mean (LC) (ppm) & log Jitter (LC) & log (sigma) (GP) & log (rho) (GP) \\
        \hline
       TOI-271 & $0.927^{+0.009}_{-0.010}$ & $16.1\pm1.6$ & $-7.642\pm0.002$ & -- & -- \\
       TOI-510 & $0.987^{+0.014}_{-0.006}$ & $29\pm16$ & $-7.621\pm0.002$ & $-7.609\pm0.038$ & $0.622\pm0.023$ \\
       TOI-641 & $0.766^{+0.039}_{-0.045}$ & $22.8\pm4.9$ & $-8.204\pm0.016$ (sectors 6, 7), $-6.805\pm0.005$ (sector 33)  & -- & -- \\
       TOI-697 & $0.21^{+0.15}_{-0.14}$ & $12.0\pm19.5$ & $-7.248\pm0.003$ & $-8.44\pm0.10$ & $1.353\pm0.095$ \\
       TOI-745 & $0.068^{+0.073}_{-0.047}$ & $49.8\pm13.6$ & $-6.368\pm0.002$ & $-8.280\pm0.028$ & $0.831\pm0.028$ \\
        \hline
    \end{tabular}
    \label{tab:fit_app}
\end{table*}

\bsp	
\label{lastpage}
\end{document}